\documentclass[
superscriptaddress,
longbibliography,
 amsmath,amssymb,
 aps,
reprint
]{revtex4-1}

\usepackage{color}
\usepackage{graphicx}
\usepackage{dcolumn}
\usepackage{bm}
\usepackage{gensymb}

\usepackage{float}

\begin{document}


\title{Remote epitaxial frustration stabilizes a correlated interfacial state}

\author{Taehwan Jung}
\affiliation{Materials Science and Engineering, University of Wisconsin-Madison, Madison, Wisconsin 53706, United States}

\author{Nicholas Hagopian}
\affiliation{Materials Science and Engineering, University of Wisconsin-Madison, Madison, Wisconsin 53706, United States}

\author{Quinn Campbell}
\affiliation{Quantum Computer Science Department, Sandia National Laboratories, Albuquerque, New Mexico 87158, United States}

\author{Anshu Sirohi}
\affiliation{Materials Science and Engineering, University of Wisconsin-Madison, Madison, Wisconsin 53706, United States}

\author{Chengye Dong}
\affiliation{Two-dimensional Crystal Consortium, Materials Research Institute, Pennsylvania State University, University Park, Pennsylvania 16802, United States}

\author{Zachary T. LaDuca}
\affiliation{Materials Science and Engineering, University of Wisconsin-Madison, Madison, Wisconsin 53706, United States}

\author{Tamalika Samanta}
\affiliation{Materials Science and Engineering, University of Wisconsin-Madison, Madison, Wisconsin 53706, United States}

\author{Joshua Robinson}
\affiliation{Two-dimensional Crystal Consortium, Materials Research Institute, Pennsylvania State University, University Park, Pennsylvania 16802, United States}
\affiliation{Department of Materials Science and Engineering, Pennsylvania State University, University Park, Pennsylvania 16802, United States}

\author{Paul M. Voyles}
\affiliation{Materials Science and Engineering, University of Wisconsin-Madison, Madison, Wisconsin 53706, United States}

\author{Jason K. Kawasaki}
\affiliation{Materials Science and Engineering, University of Wisconsin-Madison, Madison, Wisconsin 53706, United States}
\email{jkawasaki@wisc.edu}

\date{\today}
\begin{abstract}
Remote epitaxy exploits substrate interactions transmitted across atomically thin materials to replicate substrate crystal structure. Here we show that competition among graphene-, substrate-, and reconstruction-derived interactions can instead produce frustration, preventing selection of a unique ordered state. Using GdAuGe films on $N$-layer graphene/SiC(0001), we identify a self-limited interfacial state with broken long-range translational order at intermediate $N$, together with non-monotonic orientation selection that departs from direct substrate registry. Annealing shows that the frustrated interface emerges from an epitaxial crystal, distinguishing it from kinetically trapped disorder. First-principles calculations reveal a multi-periodic interfacial potential that provides a microscopic basis for frustration. The frustrated state further exhibits enhanced magnetic irreversibility above 300 K, with low-field response scaling with interface area rather than film volume. Remote epitaxial frustration therefore provides a route to stabilize finite-range interfacial order and associated collective properties.
\end{abstract}

\maketitle


Frustration arises when competing interactions generate multiple nearly degenerate free-energy minima, preventing selection of a unique ordered state \cite{tarjus2005frustration, frank1952supercooling}. Rather than producing random disorder, frustration can stabilize finite-range correlated states distinct from crystalline and amorphous limits, as found in spin glasses \cite{binder1986spin,sherrington1975solvable}, relaxor ferroelectrics \cite{hemberger2005relaxor,cross1987relaxor}, spin ice \cite{ramirez1999zero}, and strain glasses \cite{sarkar2005evidence}. Yet frustration has rarely been exploited in crystal growth, where epitaxy instead seeks a unique crystallographic registry corresponding to a single interfacial free-energy minimum \cite{cho1975molecular, baliga1986silicon}.

Remote epitaxy through atomically thin graphene \cite{kim2017remote,kong2018polarity} or amorphous carbon \cite{Kim2023_highthroughput,jia2025long} interlayer provides a natural platform to realize this missing regime. 
Previous studies have focused primarily on the limit in which substrate-derived interactions transmitted \cite{kim2017remote, kong2018polarity, liu2025remote, jiang2019carrier, yoon2022freestanding,chang2023remote,jia2025long, dai2022highly} or mediated \cite{wang2023modulation} through graphene dominate epitaxial selection.
Here we consider the qualitatively different case in which graphene-, substrate-, and reconstruction-derived interactions with competing spatial periodicities \cite{laduca2025transparent, kawasaki2025model, kim2021impact} become comparable, giving rise to multiple nearly degenerate structural configurations. We term this regime remote epitaxial frustration (Fig. \ref{fig:concept}).

Here we demonstrate remote epitaxial frustration in GdAuGe films grown on $N$-layer graphene/6H-SiC(0001). At intermediate $N$, scanning transmission electron microscopy reveals a self-limited interfacial state with finite-range structural order that emerges upon further annealing an epitaxial crystal. Orientation selection is also non-monotonic, switching from R0 for $N=0$, $N=2$, and H-intercalated graphene to R30 orientation for $N=1$. First-principles calculations reveal competing substrate-, graphene-, and reconstruction-derived periodic interactions at intermediate $N$, providing a microscopic basis for the frustrated landscape. The frustrated sample further exhibits strongly enhanced magnetic irreversibility persisting above room temperature, linking the finite-range interfacial state to a distinct collective magnetic response.

Our findings also provide more discriminating evidence for remote substrate influence. Film registry to the substrate in the presence of graphene can also arise from direct nucleation through graphene openings \cite{manzo2022pinhole, du2022controlling, jang2023thru, zulqurnain2022defect, yan2022situ, laduca2025transparent}, or from serial epitaxy \cite{laduca2025transparent} in which the film couples directly to graphene that is itself registered to the substrate. In contrast, the finite-range interfacial state and non-monotonic orientation selection observed here emerge only at intermediate $N$ and are absent from both the bare-substrate and substrate-decoupled limits. These non-additive responses therefore provide a mechanism-discriminating signature of remotely transmitted interactions.

\section*{Results}

\begin{figure}[h]
    \centering
    \includegraphics[width=1\linewidth]{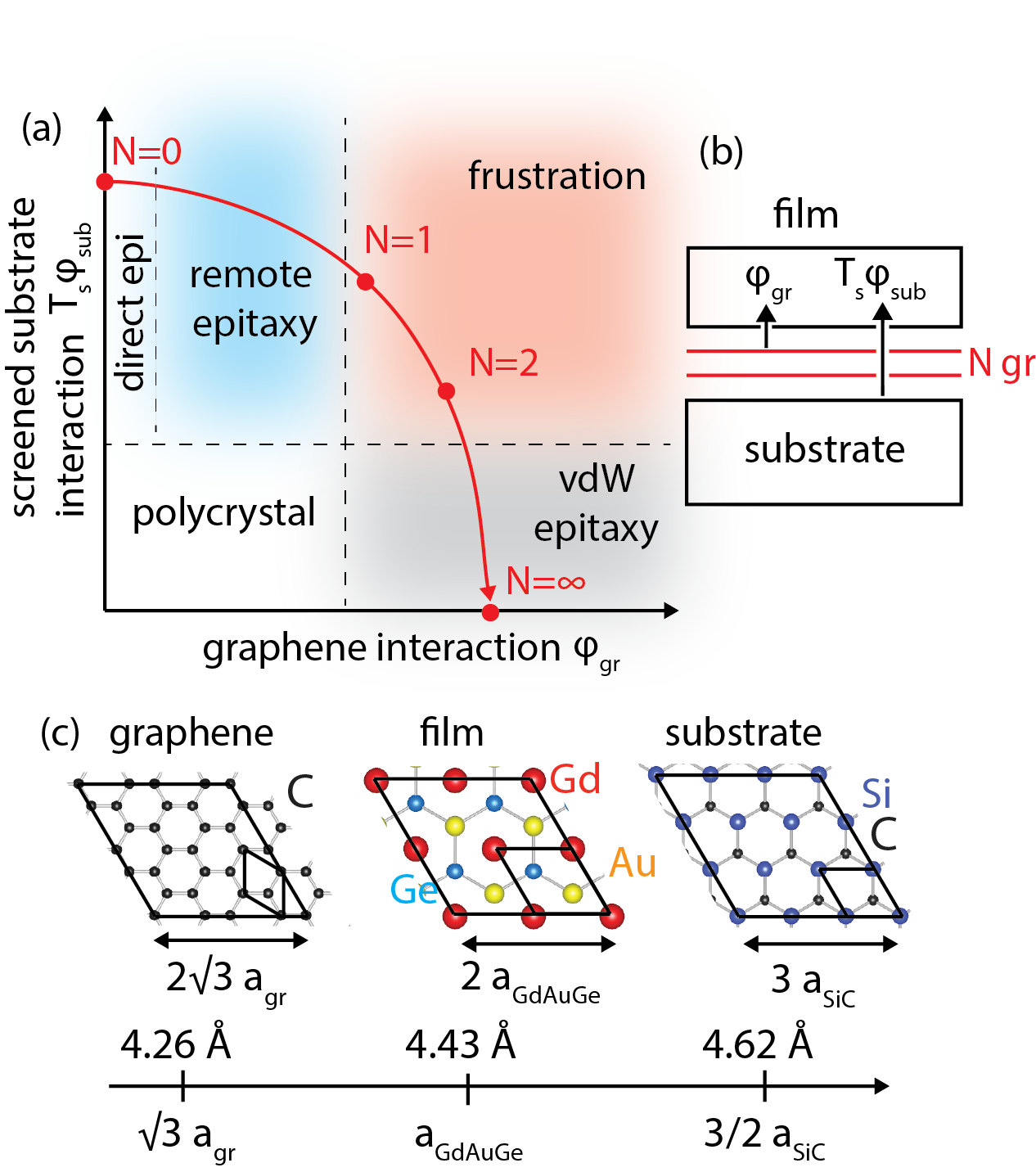}
    \caption{\textbf{Concept of remote frustration.} (a) Schematic regimes of growth, traversed via number of graphene layers $N$. The reconstruction interaction $\varphi_{rec}$ is omitted for simplicity. (b) Film/graphene/substrate heterostructure. (c) Selection of GdAuGe film ($P6_3 mc$) on graphene/SiC.}
    \label{fig:concept}
\end{figure}

Fig. \ref{fig:concept}(a,b) illustrates the concept for remote epitaxial frustration. A film on a graphene-covered substrate experiences a direct interaction to graphene $\varphi_{gr}$, a screened remote interaction to the underlying substrate $\varphi_{sub}$, and an interaction with the graphene/substrate interfacial reconstruction $\varphi_{rec}$, each with a distinct spatial periodicity. We capture this phenomenology with the expression
\begin{equation}
    \varphi_{total} = \varphi_{gr} + T_s (\varphi_{sub} + \varphi_{rec}),
\end{equation} 
where $T_s=0$ to 1 describes the screened transmission through graphene and depends on graphene thickness and carrier density \cite{kawasaki2025model,laduca2025transparent}.
When $\varphi_{gr}$ dominates, the system approaches the graphene-dominated (van der Waals) regime. When the transmitted substrate interaction dominates, it approaches the substrate-dominated (remote) regime. Between these limits, we hypothesize that graphene-, substrate-, and reconstruction-derived interactions of comparable magnitude can compete, preventing selection of a unique epitaxial registry and stabilizing finite-range rather than long-range translational order. Graphene thickness $N$, or carrier density $n$, therefore provides a means to tune across these regimes (Fig. \ref{fig:concept}(a)).

\begin{figure*}[t]
    \centering
    \includegraphics[width=.95\textwidth]{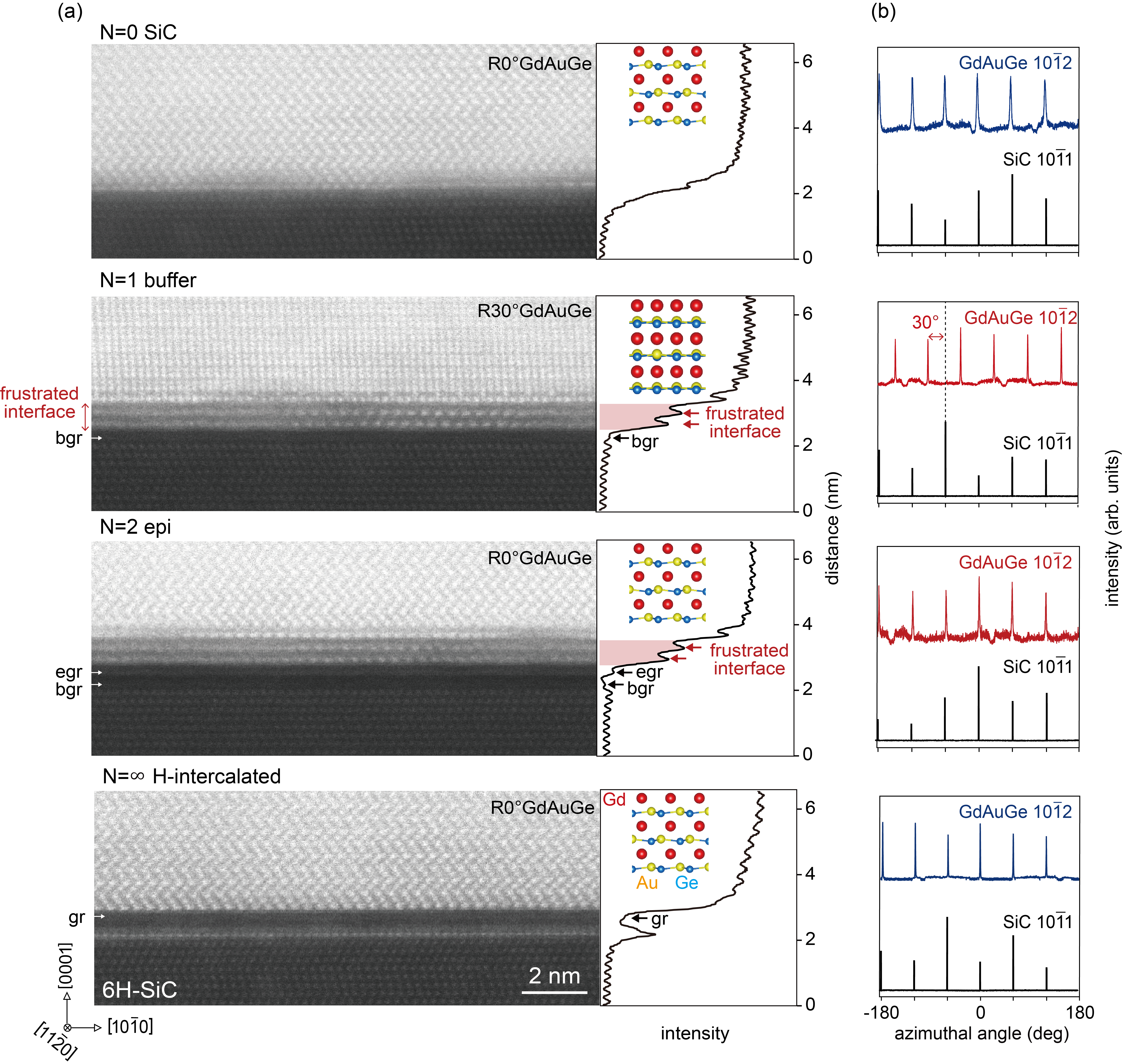}
   \caption{\textbf{Interface with broken long-range translational order and graphene-dependent orientation selection.} (a) Cross-sectional HAADF-STEM showing a self-limited, 2-3 layer thick interface with broken long-range translation on epitaxial graphene (egr) and buffer graphene (bgr). Gd (red), Au (yellow), and Ge (blue). (b) X-ray azimuthal scans showing non-monotonic orientation selection: GdAuGe on buffer graphene adopts R30 orientation, whereas all other films adopt R0.
    }
    \label{fig:tem}
\end{figure*}

\subsection*{Remote frustration and orientation selection}

We synthesized GdAuGe films by molecular beam epitaxy on bare SiC ($N=0$), buffer graphene ($N=1$), epitaxial graphene ($N=2$), and H-intercalated graphene/6H-SiC(0001), using a room-temperature seed followed by annealing to $450\degree$C \cite{laduca2024cold} (Methods).
Buffer and epitaxial graphene were synthesized directly on SiC via Si sublimation, while H-intercalation decouples buffer graphene from the substrate and provides the van der Waals limit \cite{sforzini2015approaching} (Supplementary Fig. \ref{supp:graphene}).
X-ray photoemission spectroscopy confirms the expected \cite{conrad2017structure} mixed $sp^2/sp^3$ character of buffer graphene and $sp^2$ character of epitaxial graphene/SiC (Supplementary Fig. \ref{supp:xps_thick}).
X-ray diffraction and atomic force microscopy confirm phase-pure epitaxial films with sharp thickness fringes and smooth atomic step-terrace surfaces (Supplementary Figs. \ref{supp:xrd}, \ref{supp:afm}).
The limiting $N=0$ and H-intercalated cases form crystalline interfaces by high-angle annular dark-field scanning transmission electron microscopy (HAADF-STEM, Fig. \ref{fig:tem}(a)) and exhibit the same R0 ($\mathbf{a}_{\mathrm{GdAuGe}} \parallel \mathbf{a}_{\mathrm{SiC}}$) epitaxial alignment by azimuthal x-ray diffraction (Fig. \ref{fig:tem}(b)).


Intermediate graphene thicknesses instead produce a broken-translation interfacial state together with non-monotonic orientation selection. On $N=1$ buffer and $N=2$ epitaxial graphene/SiC, HAADF-STEM reveals a self-limited, 2--3 atomic layer interfacial state with local order but broken long-range lateral translation (Fig. \ref{fig:tem}a; Supplementary Fig. \ref{supp:TEM_wide}). Its occurrence on both $sp^2$ epitaxial graphene and mixed $sp^2/sp^3$ buffer graphene, but absence on $sp^2$ H-intercalated graphene, argues against local graphene bonding as its origin. Moreover, Gd or Au intercalation is observed beneath H-intercalated graphene \cite{briggs2020atomically}, but not at the $N=1$ and $N=2$ interfaces. 
Together with similar in-situ photoemission lineshapes across these surfaces (Supplementary Figs. \ref{supp:xps} and \ref{supp:xps_thick}), these observations argue against conventional interdiffusion or chemical reaction and instead point to competing graphene-, substrate-, and reconstruction-derived interactions as the origin of the interfacial state.

Orientation selection also varies non-monotonically with graphene thickness. Films on bare SiC ($N=0$), epitaxial graphene ($N=2$), and H-intercalated graphene adopt R0 alignment to SiC, whereas $N=1$ buffer graphene selects R30 ($[1 1 \bar 2 0]_{\mathrm{GdAuGe}} \parallel [1 0 \bar 1 0]_{\mathrm{SiC}}$). R30 is observed locally by HAADF-STEM (Fig. \ref{fig:tem}(a) and Supplementary Fig. \ref{supp:TEM_wide}) and over millimeter-scale areas by x-ray azimuthal diffraction (Fig. \ref{fig:tem}(b)). 
This graphene-dependent orientation selection is also observed within a single sample: across an atomic step separating adjacent buffer- and epitaxial-graphene terraces, GdAuGe switches from R30 to R0 despite identical growth conditions, directly linking orientation selection to the local graphene thickness (Supplementary Fig. \ref{supp:step}).
GdAuSb exhibits the same $R0 \rightarrow R30 \rightarrow R0$ sequence on bare SiC, buffer graphene, and epitaxial graphene, respectively (Supplementary Fig. \ref{supp:GdAuSb}), while our previous studies show GdPtSb similarly switches from R0 on bare sapphire to R30 on graphene/sapphire \cite{du2022controlling}. Thus the graphene layer dependent orientation selection is not specific to one particular film/substrate combination.

Together, the broken-translation interface and graphene-dependent orientation selection provide evidence for remote substrate influence without requiring replication of the substrate registry. In the conventional substrate-dominated limit ($T_s\varphi_{sub}\gg\varphi_{gr}$), substrate replication \cite{kim2017remote, kong2018polarity, liu2025remote, jiang2019carrier, yoon2022freestanding, chang2023remote} is intrinsically difficult to distinguish from direct nucleation through graphene openings followed by lateral growth \cite{manzo2022pinhole, du2022controlling, jang2023thru,zulqurnain2022defect,yan2022situ,laduca2025transparent}. Here, intermediate graphene coupling instead produces outcomes absent from both limiting cases: the broken-translation interface occurs only for $N=1$ and $N=2$, while buffer graphene selects R30 rather than the R0 registry of both bare SiC and substrate-decoupled graphene. The former identifies an emergent interfacial state, while the latter provides complementary evidence that the underlying substrate influences growth without simply imposing its registry. Although the edges of graphene pinholes can locally perturb film orientation \cite{chang2023remote,an2026orientation}, such effects do not readily explain the global selection of a R30 registry over millimeter scales (Fig. \ref{fig:tem}(b)).

\subsection*{Thermal evolution into the frustrated state}

\begin{figure*}
    \centering
    \includegraphics[width=1\textwidth]{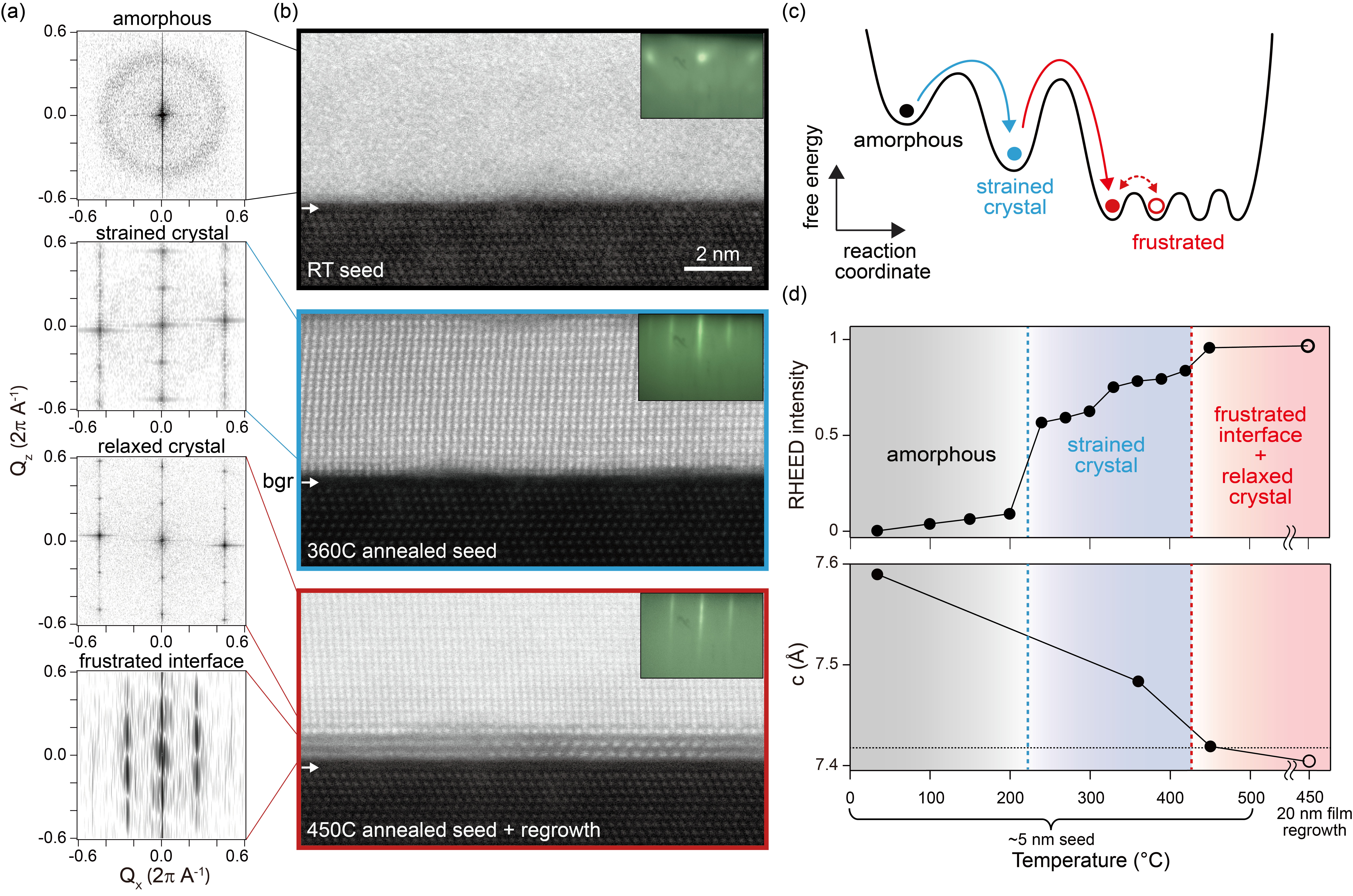}
    \caption{\textbf{Formation of the frustrated interface.} (a) Fast Fourier Transforms (FFT) of selected film and interface regions as a function of annealing. (b) STEM images of GdAuGe seed layer deposited at room temperature on buffer graphene/SiC as a function of subsequent annealing. The buffer graphene layer is noted by the white arrow. The corresponding RHEED patterns along the $\langle 10\bar1 0 \rangle_{SiC}$ are shown in the insets. (c) Schematic free-energy landscape consistent with the observed anneal evolution. (d) Normalized RHEED intensity ($I_{diffracted}/I_{background}$) and out of plane lattice parameter from x-ray diffraction, as function of anneal temperature. The bulk lattice parameter is marked by the dotted line. See Supplementary Fig. \ref{supp:anneal_xrd} for the raw XRD evolution.
    }
    \label{fig:anneal}
\end{figure*}

To determine how the broken-translation interface develops, we tracked the structural evolution of a $\sim5$ nm thick GdAuGe seed layer on $N=1$ buffer graphene/SiC during annealing (Fig. \ref{fig:anneal}). The as-grown room-temperature seed is amorphous, as indicated by the diffuse reflection high energy electron diffraction (RHEED) pattern, diffuse ring pattern in the fast Fourier transform (FFT) of the STEM image, and the absence of Bragg x-ray diffraction peaks (Supplementary Fig. \ref{supp:anneal_xrd}). Annealing above $200\degree$C produces a strained R30 crystal as measured by x-ray diffraction (Supplementary Fig. \ref{supp:anneal_xrd}) with a crystalline graphene interface observed by STEM (Fig. \ref{fig:anneal}). Only upon further annealing to $450\degree$C does the lattice relax while simultaneously generating the self-limited broken-translation interfacial layer. A reconstructed layer with $3\times$ lateral periodicity forms between the broken-translation interface and relaxed crystal, potentially accommodating the lattice mismatch. 

Our experiments show that the finite-range interfacial state is not kinetically trapped disorder from incomplete crystallization. Upon annealing, the interface evolves from an amorphous seed, to an R30 epitaxial crystal, and finally to the broken-translation state with only finite-range translational order (Fig. \ref{fig:anneal}(c)). The intermediate formation of a crystalline interface before the emergence of finite-range order demonstrates that the latter is not inherited from poorly aligned nuclei or incomplete crystallization. Instead, the reversal from increasing to decreasing translational order suggests that additional competing interfacial configurations become accessible at elevated temperature. Importantly, direct growth at $450^\circ$C converges to the same broken-translation interface and R30 overlayer orientation as the room-temperature seed-plus-anneal pathway (Supplementary Fig. \ref{supp:history}). The convergence of distinct growth histories to the same final state further argues against a pathway-specific kinetic artifact. Although direct high-temperature growth produces greater intercalation beneath graphene, the broken-translation interface occurs both with and without appreciable intercalation, indicating that intercalation is not required for its formation.

Despite the loss of translational order at the interface, the crystalline film above remains exceptionally well ordered, with an x-ray rocking-curve width of only 6 arc sec (Supplementary Fig. \ref{supp:xrd}e). The broken-translation layer exhibits a lateral correlation length $\xi=2\pi/\Delta Q_x=30$ \AA, extracted from the STEM FFT peak width $\Delta Q_x$, intermediate between that of the amorphous film ($\xi=10$ \AA) and relaxed crystal ($\xi>100$ \AA; Supplementary Fig. \ref{supp:fft_width}). This finite correlation length distinguishes the interface from conventional strain relaxation and domain-matching epitaxy \cite{matthews1974defects, narayan2003domain,sato2025superconductivity}, which accommodate mismatch through misfit dislocations while retaining crystalline registry, and from conventional interfacial reconstructions that introduce new long-range periodicities.

\subsection*{Magnetic response of the frustrated state}

\begin{figure}[h]
    \centering
    \includegraphics[width=1\linewidth]{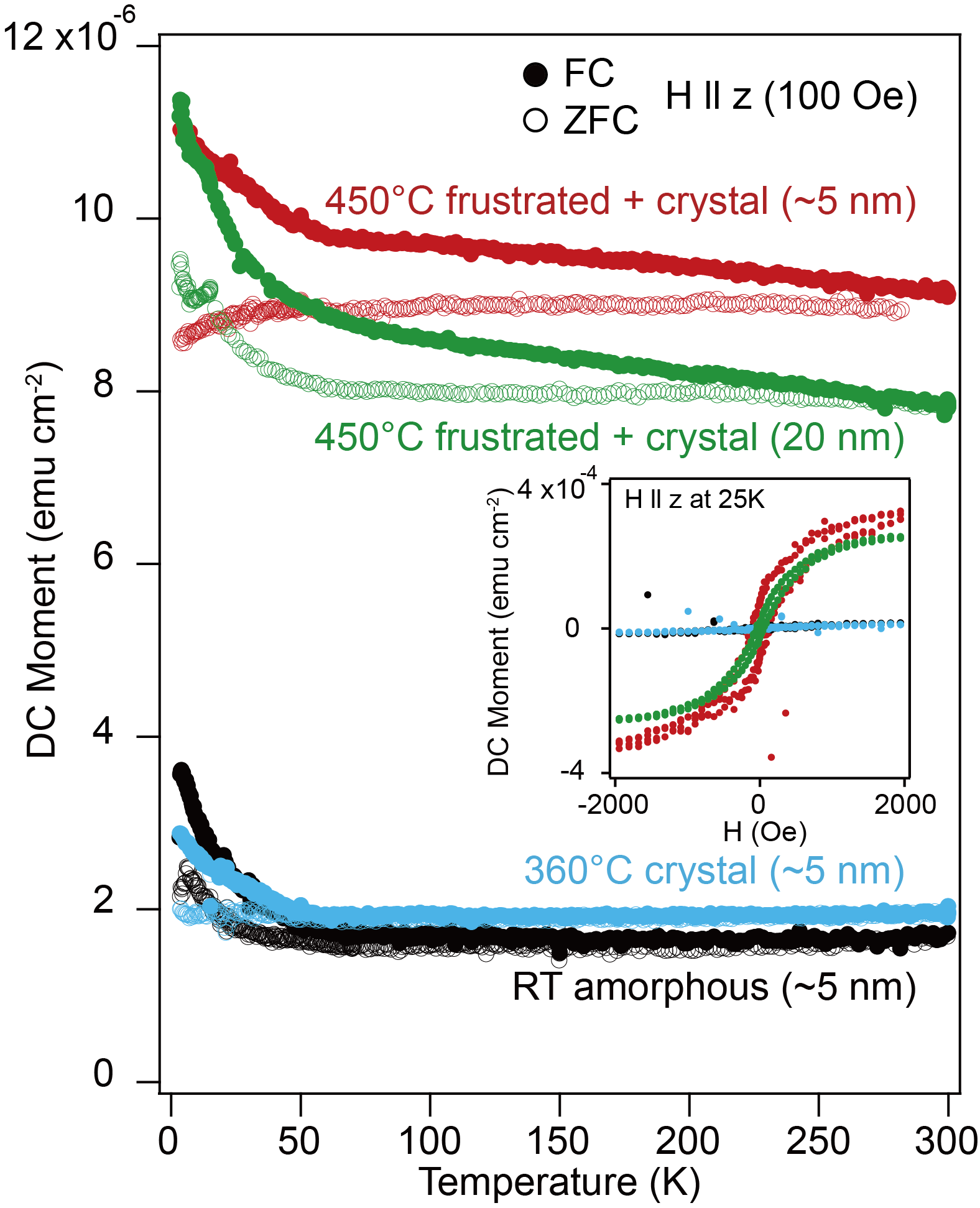}
    \caption{\textbf{Interface-driven magnetic response of the frustrated state.} Field cooled (FC) and zero field cooled (ZFC) magnetic moment versus temperature showing a bifurcation that persists above room temperature for samples with frustrated interface on buffer graphene/SiC (red, green), and weaker magnetic response for fully crystalline (blue) and amorphous (black) samples annealed to lower temperatures. Inset shows $M(H)$ at 25 K.}
    \label{fig:magnet}
\end{figure}

The frustrated structural state is accompanied by a strongly enhanced magnetic response (Fig. \ref{fig:magnet}). Despite a $4\times$ difference in film thickness, the $\sim5$ and 20 nm samples on buffer graphene exhibit comparable area-normalized moments and pronounced ZFC–FC irreversibility persisting to 300 K, indicating that the enhanced response scales with interface area rather than GdAuGe volume. Similar behavior occurs for frustrated interfaces on epitaxial graphene (Supplementary Fig. \ref{supp:magnet20nm}), whereas crystalline GdAuGe on bare SiC and amorphous or strained-crystalline controls on graphene/SiC show much smaller moments and weaker irreversibility (Fig. \ref{fig:magnet}). Consistently, $M(H)$ at 25 K is hysteretic for the frustrated samples but approximately linear for the amorphous and crystalline controls. 
Although the microscopic magnetic state is not yet distinguished among competing scenarios (e.g., cluster spin glass, superparamagnetism, or short-range exchange disorder), finite-range variations in coordination and strain may generate a heterogeneous exchange landscape, with strain gradients potentially contributing through flexomagnetic coupling \cite{laduca2024cold,samanta2026strain}. The magnetic response therefore establishes that the frustrated interface is not simply structurally disordered, but produces collective behavior distinct from both crystalline and amorphous limits.

\subsection*{Microscopic origin of competing interactions}

\begin{figure*}[t]
    \centering
    \includegraphics[width=0.9\textwidth]{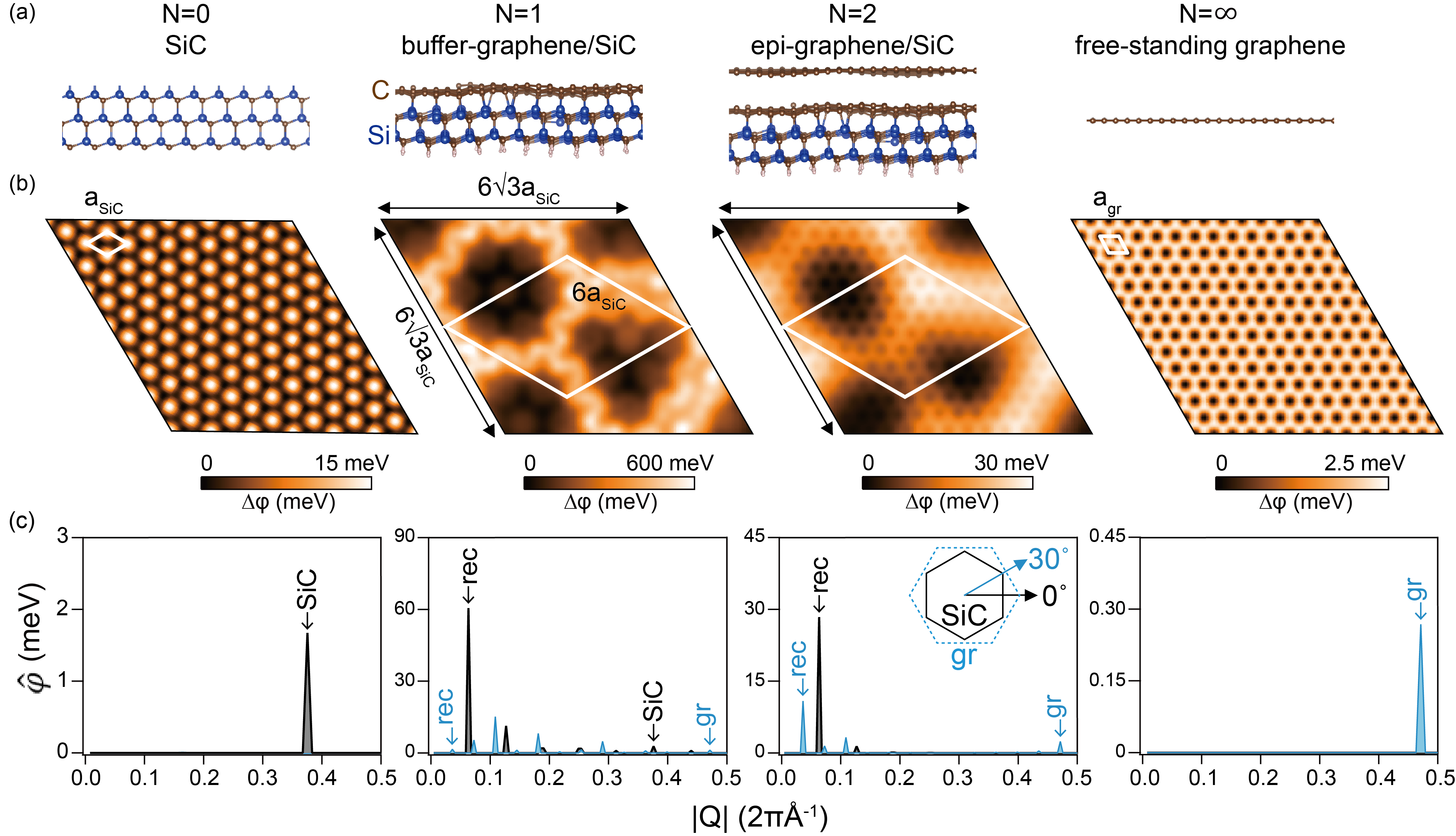}
    \caption{\textbf{Competing graphene, substrate, and reconstruction interactions above $N$-graphene/SiC.} (a) Models for 6H-SiC (0001), buffer graphene / SiC, epitaxial graphene / SiC, and isolated graphene. (b) Calculated electrostatic potential $\varphi$ at 3 \AA\ above each surface. (c) Linecuts through the Fourier transform $\hat{\varphi}(\vec{Q})$, showing single frequencies for bare SiC and freestanding graphene, and multiple peaks above buffer and epitaxial graphene on SiC. $Q_{6\sqrt{3}} = 0.036$ ($2\pi$/\AA) and $Q_6 = 0.063$ ($2\pi$/\AA) correspond to the $(6\sqrt3 \times 6\sqrt3)R30\degree$ and $(6\times6)$ reconstructed periodicities, respectively. $0\degree$ cuts (black) are for $\mathbf{Q}\parallel [11\bar 2 0]_{\mathrm{SiC}}$, $30\degree$ cuts (blue) are for $\mathbf{Q}\parallel [10 \bar 1 0]_{\mathrm{SiC}}$. 
    }
    \label{fig:potentials}
\end{figure*}

We next examine the microscopic interactions underlying the frustrated interfacial state and non-monotonic orientation selection. The annealing evolution and finite-range structural order suggest an interfacial free-energy landscape with multiple competing configurations (Fig. \ref{fig:anneal}(c)). We calculate the electrostatic potential above $N$-layer graphene/SiC using density functional theory and decompose its spatial Fourier components $\hat{\varphi}(Q)$ (Fig. \ref{fig:potentials}, Methods). The limiting cases each exhibit a single dominant spatial periodicity: $Q_{sub}$ for bare SiC ($N=0$) and $Q_{gr}$ for isolated graphene. In contrast, buffer and epitaxial graphene/SiC exhibit graphene-, substrate-, and reconstruction-derived periodicities ($Q_{6}$ and $Q_{6\sqrt{3}}$), with multiple components of comparable amplitude and no single periodicity dominating. Adding a second graphene layer suppresses most substrate- and reconstruction-derived Fourier amplitudes by a factor of two, consistent with increased screening. Thus, graphene thickness changes both the strength of the transmitted interaction amd the spectrum of spatial interactions defining the interfacial landscape. Multiple graphene- and substrate-derived contributions have appeared in previous calculations of remote epitaxial potentials \cite{dai2022highly,kim2021impact}; here, the additional reconstruction-derived components and their comparable amplitudes produce the competing multi-periodic landscape associated with the frustrated regime.

To determine whether these multiple spatial interactions translate into competing structural configurations, we calculate the formation energy $E_{\mathrm{form}}(\theta,x,y)$ of finite GdAuGe islands (Supplementary Figs. \ref{supp:rotation_model}, \ref{supp:rotation_small}, \ref{supp:rotation_gr}). Bare SiC yields a robust minimum near $\theta=0^\circ$, consistent with the observed R0 alignment, whereas buffer and epitaxial graphene exhibit several shallow, registry-dependent rotational minima. 
Because their relative energies depend on island size and edge termination (Supplementary Fig. \ref{supp:rotation_gr}), we do not interpret these calculations as quantitative predictions of the preferred registry. Rather, they demonstrate that the multi-periodic potential identified in $\hat{\varphi}(Q)$ can generate multiple competing structural configurations. Notably, finite-range interfacial order occurs only for the intermediate graphene thicknesses for which the calculations identify multiple competing spatial interactions, whereas the experimental $N=0$ bare SiC and H-intercalated graphene limits form crystalline interfaces. The conjunction of three observations -- the broken translation interface with self-limited thickness, its restriction to intermediate graphene coupling, and the calculated competition among spatial interactions in this same regime -- provides the basis for identifying the interfacial state as frustrated rather than kinetically disordered.

\section*{Discussion}

Our results establish remote epitaxial frustration as a distinct growth regime in which competing graphene, substrate, and reconstruction-derived interactions stabilize finite- rather than long-range interfacial order, distinct from kinetically trapped disorder or incomplete crystallization. The accompanying strongly enhanced magnetic irreversibility demonstrates that this frustration is not merely structural, but can generate collective behavior distinct from both crystalline and amorphous limits.

Remote epitaxial frustration provides a more discriminating signature of remotely transmitted substrate interactions. Substrate-aligned growth is not mechanism specific, as it can arise from nucleation through graphene openings \cite{manzo2022pinhole, du2022controlling, jang2023thru,zulqurnain2022defect,yan2022situ,laduca2025transparent} or via serial epitaxy on substrate-registered graphene \cite{laduca2025transparent}. In contrast, competing interactions produce non-additive structural outcomes absent from the bare-substrate and decoupled-graphene limits, either shifting the preferred registry or suppressing long-range order altogether, as manifested by the $R0 \rightarrow R30 \rightarrow R0$ orientation sequence and finite-range interfacial state.

More broadly, remote epitaxial frustration expands epitaxy from selecting crystallographic registry to engineering emergent properties through the interfacial free-energy landscape. The enhanced magnetism of frustrated GdAuGe illustrates this opportunity: interaction competition can stabilize collective behavior absent from both crystalline and amorphous limits. 
In the present system, graphene layer number $N$ provides a discrete means to tune across these regimes; carrier density $n$ offers a possible continuous and in situ knob to control the frustrated state and its collective response.
Remote epitaxial frustration therefore establishes interaction competition as a means to jointly engineer structural correlations and emergent properties at buried interfaces.

\section*{Acknowledgments}

\section*{Funding}

This work was primarily supported by the U.S. Department of Energy, Office of Science, Basic Energy Sciences, under award no. DE-SC0023958 (GdAuGe synthesis and XPS by T.J., A.S., J.K.K). Additional support came from the NSF through the University of Wisconsin Materials Research Science and Engineering Center under Grant No. NSF DMR-2309000 (magnetometry and structural characterization by N.H., T.S., J.K.K., and PMV) and the Air Force Office of Scientific Research FA9550-21-0127 (film characterization by Z.L., J.K.K.). J.K.K. acknowledges support from the Gordon and Betty Moore Foundation (DOI:10.37807/GBMF13808) for data analysis. We acknowledge the use of facilities and instrumentation in the Wisconsin Center for Nanoscale Technology. This Center is partially supported by the Wisconsin Materials Research Science and Engineering Center (NSF DMR-2309000) and by the University of Wisconsin–Madison.

Graphene on SiC samples for this publication were provided by The Pennsylvania State University Two-Dimensional Crystal Consortium – Materials Innovation Platform (2DCC-MIP) which is supported by NSF cooperative agreement DMR-2039351 (C.D. and J.R.).

QTC is supported by the Laboratory Directed Research and Development (LDRD) program at Sandia National Laboratories under project 233271.
This work was performed, in part, at the Center for Integrated Nanotechnologies, an Office of Science User Facility operated for the U.S. Department of Energy (DOE) Office of Science.
Sandia National Laboratories is a multi-mission laboratory managed and operated by National Technology \& Engineering Solutions of Sandia, LLC (NTESS), a wholly owned subsidiary of Honeywell International Inc., for the U.S. Department of Energy’s National Nuclear Security Administration (DOE/NNSA) under contract DE-NA0003525. This written work is authored by an employee of NTESS. The employee, not NTESS, owns the right, title and interest in and to the written work and is responsible for its contents. Any subjective views or opinions that might be expressed in the written work do not necessarily represent the views of the U.S. Government. The publisher acknowledges that the U.S. Government retains a non-exclusive, paid-up, irrevocable, world-wide license to publish or reproduce the published form of this written work or allow others to do so, for U.S. Government purposes. The DOE will provide public access to results of federally sponsored research in accordance with the DOE Public Access Plan.

\section*{Author contributions}

TJ, QC, and JKK conceived and planned the study. TJ, ZTL, and TS synthesized GdAuGe films, supervised by JKK. NH performed STEM measurements, supervised by PMV. QC performed first principles calculations. AS performed XPS studies. CD and JR synthesized graphene on SiC. TJ performed x-ray diffraction and magnetometry measurements. TJ and JKK wrote the manuscript with contributions from all authors.

\section*{Competing interests}

The authors declare no competing interests

\section*{Additional information}
Supplementary information is available for this paper.
Correspondence and requests for materials should be addressed to Jason K. Kawasaki.

\section*{Methods} \label{sec:methods}

\subsubsection*{Graphene growth on SiC and characterization.} Buffer ($N=1$) and epitaxial ($N=2$) graphene are grown by Si sublimation from 6H-SiC (0001) (Coherent Inc.) via thermal decomposition to avoid transfer-induced graphene contaminants. Buffer $N=1$ graphene is grown just beyond complete monolayer coverage to minimize the impact of exposed areas to the substrate \cite{manzo2022pinhole} . This produces an extra layer of $N=2$ epitaxial graphene localized the step edge (Supplementary Fig. \ref{supp:graphene}(d,g)). We use hydrogen-intercalated graphene to approximate the limit of isolated graphene. H-intercalated graphene is also called quasi-freestanding graphene in the literature. 

The SiC substrates were first annealed at $1400 \degree$C and 700 Torr in a 10 \% $H_{2}$/Ar mixture for 30 min to remove surface contamination. Subsequently, the samples were annealed in pure Ar at $1600 \degree$C and 700 Torr for 30 min to form the buffer layer, and at $1800 \degree$C and 700 Torr for 20 min to produce monolayer epitaxial graphene, respectively. 
Quasi-freestanding monolayer graphene was prepared by hydrogen intercalation of buffer layer at $700 \degree$C and 600 Torr in pure $H_{2}$ for 30 min.

Raman spectroscopy is performed in a Horiba LabRam Raman system with a 532 nm laser and a 600 grooves cm$^{-1}$ grating. The spectra are obtained with a power of 6 mW and an integration time of 15 s. Atomic force microscopy (AFM) is performed using Bruker Dimension Icon instrument equipped with a ScanAsyst-Air ($k = 0.4$ N/m) tip in tapping mode.

\subsubsection*{MBE growth of GdAuGe.} GdAuGe films were synthesized by molecular beam epitaxy (MBE) on $N$-layer graphene/6H-SiC(0001) substrates using a room temperature seed plus anneal approach to improve surface wetting \cite{laduca2024cold}. Briefly, a $\sim$5 nm thick GdAuGe seed is deposited at room temperature via co-evaporation from elemental Gd, Au, and Ge sources. The seed is then annealed to $450\degree$C, followed by continued growth at $450\degree$C to achieve a total film thickness of approximately 20 nm  \cite{laduca2024cold}. Elemental fluxes were measured immediately prior to growth using a quartz crystal microbalance and calibrated to their absolute fluxes via Rutherford backscattering spectrometry (RBS) on separate calibration samples.

\subsubsection*{X-ray diffraction.} Symmetric $\omega-2\theta$, reciprocal space maps, azimuthal scans, and rocking curve measurements of GdAuGe films were performed using a Malvern Panalytical Empyrean X-ray diffractometer equipped with a four-circle goniometer. All measurements were conducted using Cu $K\alpha$ radiation ($\lambda = 1.5406 $ \AA) with a 4-bounce Ge $220$ monochromator. High resolution rocking curve measurements utilized an additional 3-bounce Ge $220$ monochromator on the analyzer side.

\subsubsection*{Transmission electron microscopy.} Cross-section lift-outs were prepared with a 30 kV Xe plasma focused ion beam (PFIB, FEI Helios) or a 30 kV Ga FIB (Zeiss Auriga). Thinning of the cross section to electron transparency was performed with the Ga FIB, with rough shaping performed at 30 kV, stepping down in voltage until a final polishing step at 2 kV. As needed, an Ar ion mill (Fischione 1040 Nanomill) was operated at 800 eV to thin cross-sections further and remove FIB surface damage and artifacts. Samples were plasma cleaned with air intake gas at 18 W for 5 min in an IBSS group MCA to remove contamination prior to STEM characterization.

Atomic resolution imaging was performed with a CEOS probe-corrected Thermo-Fisher Titan scanning transmission electron microscope (STEM). The STEM was operated at 200 kV with spot size 10 and a 70 um/23.3 mrad aperture, corresponding to a beam current of 22.3 pA. Imaging was conducted with a high angle annular dark field (HAADF) detector at a camera length of 160 mm corresponding to detection angles between 54 to 270 mrad. 

\subsubsection*{in situ XPS.} Samples were transferred from the MBE to an XPS (x-ray photoemission spectroscopy) system via an ultrahigh vacuum manifold. XPS measurements were performed at room temperature using an Omicron EA125 analyzer and an un-monochromated Mg $K\alpha$ x-ray source. The Fermi energy was calibrated using a gold film that is in electrical contact with the sample. At a typical pass energy of 20 eV, the instrumental energy resolution is 1.0 eV, calibrated by the width of the Gd 4f.

\subsubsection*{SQUID magnetometry.} Magnetometry was performed using a Quantum Design MPMS SQUID magnetometer, with field oriented out of plane. For zero field cooling (ZFC), samples are cooled from room temperature to 2 K with zero applied field, then a field of $H=100$ Oe is applied and the magnetization $M$ is measured continuously with increasing temperature. For field cooling (FC), samples are cooled from room temperature with an applied field of $H=100$ Oe to 2 K, then measured in the warming direction with the same applied magnetic field. For all curves we subtract the background measurement of a bare SiC substrate.

\subsubsection*{First-principles calculations.} We created a SiC $(6\sqrt{3}\times6\sqrt{3})R30\degree$ slab, and bulk SiC slab model where the bottom layers of the substrate were terminated with passivating hydrogen and the top layers were either exposed to vacuum or connected with a buffer graphene or epitaxial graphene layer. 
In our pure graphene slab model, we only include one layer of graphene with no passivation.
20 \AA\ of vacuum region was placed between the ends to avoid unphysical interaction between the two surfaces. 

Electronic structure calculations are done using the QUANTUM ESPRESSO package \cite{giannozzi2009quantum}. We use norm-conserving pseudopotentials from the PseudoDojo repository \cite{van2018pseudodojo} and the Perdew-Burke-Ernzerhof exchange-correlation functional \cite{perdew1996generalized}. We use kinetic energy cutoffs of 50 Ry and 400 Ry for the plane wave basis sets used to describe the Kohn-Sham orbitals and charge density, respectively. We use a 2$\times$2$\times$1 Monkhorst-Pack grid \cite{monkhorst1976special} to sample the Brillouin zone in our calculations.

\subsubsection*{Use of generative artificial intelligence.} OpenAI ChatGPT was used for language editing and to stress-test scientific arguments and alternative interpretations during manuscript preparation. All suggestions were independently evaluated by the authors, who retain full responsibility for the scientific content and conclusions.

\section*{Data Availability}

The data that support the findings of this article are openly available at doi:XX (to be provided upon acceptance)

\section*{Code availability}
No custom code central to the conclusions of this study was developed.

\newpage \clearpage \onecolumngrid 
\setcounter{page}{1}
\setcounter{figure}{0}
\renewcommand{\thefigure}{S-\arabic{figure}}
\renewcommand{\thetable}{S-\arabic{table}}
\section*{Supplement}

\subsection*{Graphene characterization.} 
As shown in Supplementary Fig. \ref{supp:graphene}(a), three representative phonon bands at $\sim$1390, $\sim$1495, and $\sim$1650 $cm^{-1}$ indicate the formation of the buffer layer \cite{wundrack2019probing}. The AFM height image (Supplementary Fig. \ref{supp:graphene}(d)) and adhesion force map (Supplementary Fig. \ref{supp:graphene}(g)) further confirm the formation of a uniform buffer layer across the terrace, except for the growth of EG ribbons at the terrace edge. After hydrogen atoms intercalated underneath the buffer layer, the buffer is converted into quasi-freestanding monolayer graphene, as evidenced by the disappearance of buffer-related features between 1200 – 1700 $cm^{-1}$ and the emergence of the graphene D peak ($\sim$1340 $cm^{-1}$), G peak ($\sim$1585 $cm^{-1}$), and 2D peak ($\sim$2665 $cm^{-1}$) \cite{sakakibara2022microscopic}.  The reduced contrast between regions near and far from the terrace edge in the adhesion-force map (Supplementary Fig. \ref{supp:graphene}(e)) also indicates effective decoupling of the buffer layer. The co-existence of broad peak spanning 1200 to 1600 $cm^{-1}$, associated with the existence of buffer layer, a G peak at $\sim$1596 $cm^{-1}$ and a 2D peak at $\sim$2720 $cm^{-1}$, which can be fitted with a single Lorentz peak, confirms the formation of an epitaxial graphene layer. The AFM height (Supplementary Fig. \ref{supp:graphene}(c)) and adhesion force mapping (Supplementary Fig. \ref{supp:graphene}(f)) images further show uniform epitaxial graphene coverage across the terrace, except for the presence of thicker graphene at the terrace edges.

\begin{figure}[h]
    \centering
    \includegraphics[width=1\linewidth]{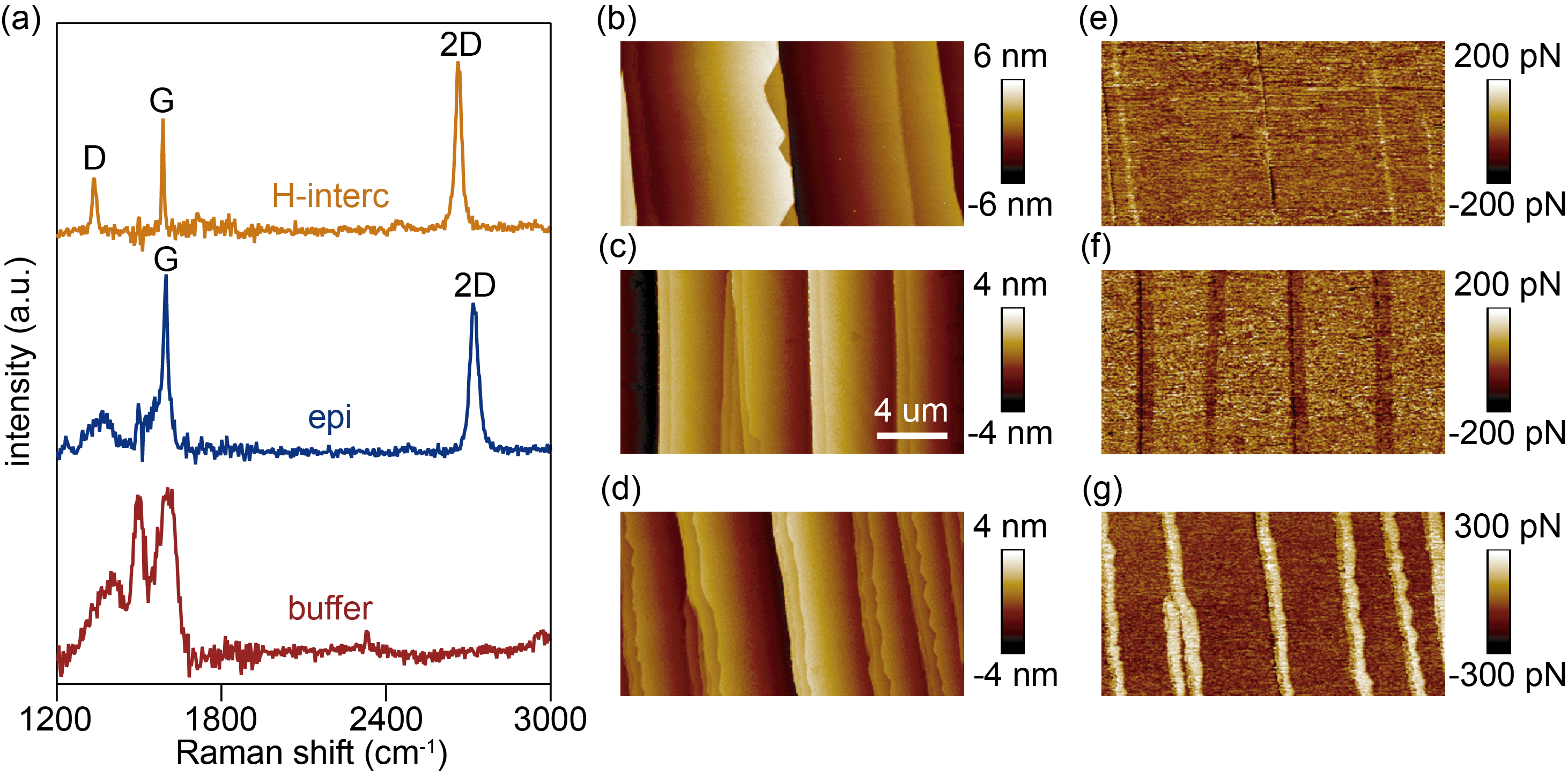}
    \caption{\textbf{Characterization of graphene/SiC substrates.} (a) Raman spectra of buffer graphene/SiC (buffer), epitaxial graphene/SiC (epi), and hydrogen-intercalated buffer graphene (H-interc). (b–d) AFM height and (e–g) adhesion-force maps of H-interc, epi, and buffer surfaces, respectively. Epitaxial-graphene ribbons/thicker graphene occur preferentially at SiC terrace edges.}
    \label{supp:graphene}
\end{figure}

\begin{figure}[h]
    \centering
    \includegraphics[width=0.8\linewidth]{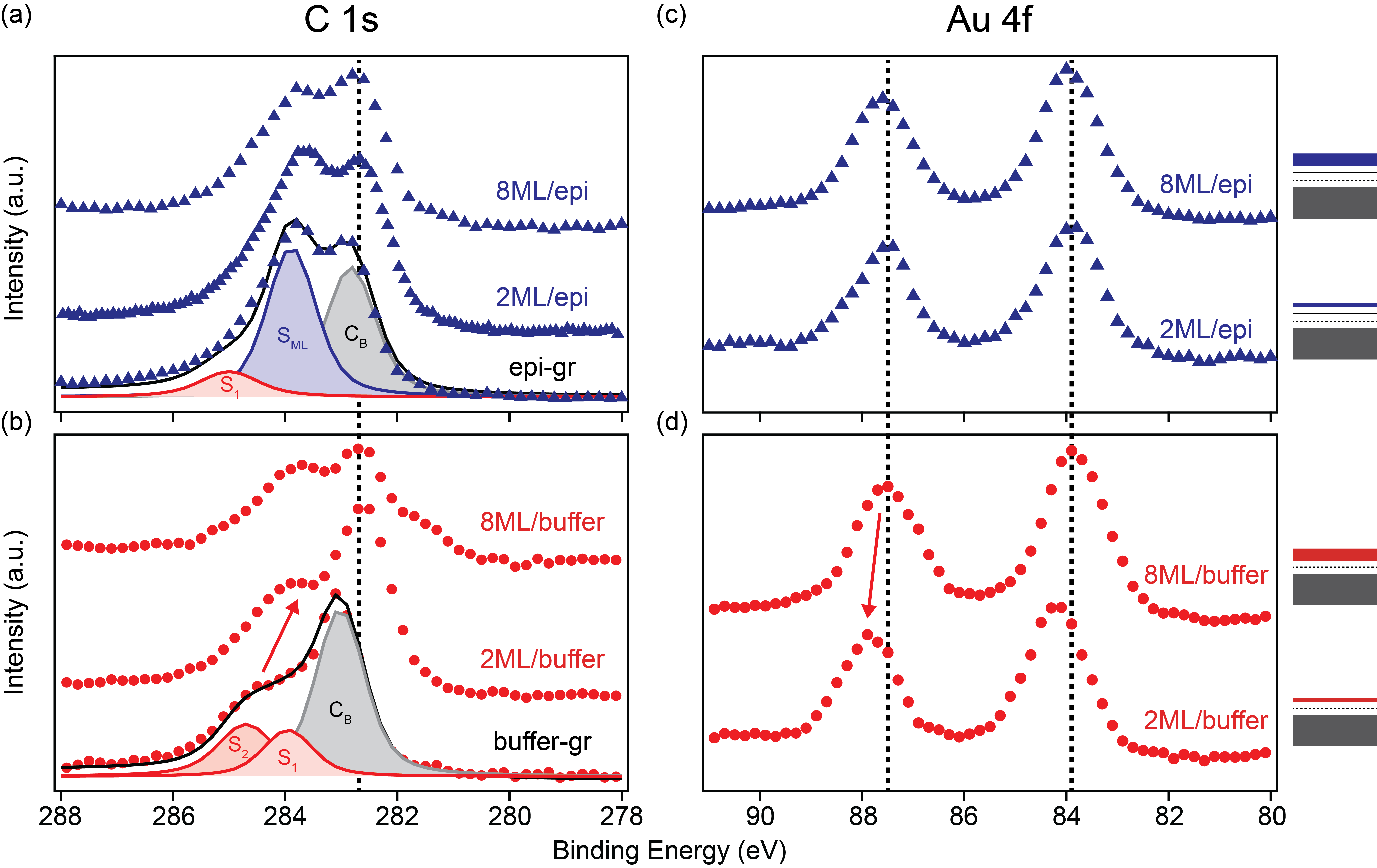}
    \caption{\textbf{Chemical-state evolution during initial GdAuGe deposition on epitaxial and buffer graphene/SiC, measured by in-situ XPS.}
    (a,b) C $1s$ for varying GdAuGe film thickness on epitaxial and buffer graphene/SiC. Shaded fit curves decompose the carbon hybridization components of the buffer and epitaxial graphene surfaces before GdAuGe growth, following Ref. \cite{conrad2017structure}: $C_B$ corresponds to bulk Si-C  $S_{1,2}$ correspond to buffer graphene with mixed $sp^2/sp^3$ hybridization, and $S_{ML}$ is the $sp^2$ component of epitaxial graphene. (c,d) Au $4f$ for varying GdAuGe thickness on epitaxial and buffer graphene. For growth on buffer graphene, the 0.5 eV binding energy shift corresponds to an Au to buffer graphene charge transfer, but no observable change in the Au $4f$ line shape.}
    \label{supp:xps_thick}
\end{figure}

\begin{table}[h]
\begin{tabular}{r|c|c|c|c}
Buffer graphene & $E_B$  (eV) & $I$ (a.u.) & $\Gamma_{L}$  (eV)\\\hline
$C_B$  & 282.92 & 0.27  & 0.5 \\
$S_1$  & 283.85 & 0.075  & 0.5 \\
$S_2$  & 284.6 & 0.1  & 0.7 \\ 
\hline
\\
Epi graphene \\
\hline
$C_B$ & 283.13 & 0.2  & 0.5 \\
$S_{ML}$ & 284.19 & 0.23  & 0.5 \\
$S_1$  &  285.3 & 0.05  & 0.8 \\
\hline
\end{tabular}
\caption{XPS C $1s$ fitting parameters for buffer and epitaxial graphene on SiC, following the assignments of Refs \cite{conrad2017structure, emtsev2008interaction}. The spectra were fitted using peak positions ($E_B$), relative intensities ($I$) with mixed Gaussian ($\Gamma_{G}$) – Lorentzian ($\Gamma_{L}$) broadening after subtraction of a Shirley-type background. The instrumental Gaussian broadening $\Gamma_{G}=0.9$ eV was determined by measurements of an Au foil under fixed slit and analyzer pass energy settings.}\label{tab:my_data}
\end{table}

\begin{figure}[h]
    \centering
    \includegraphics[width=1\linewidth]{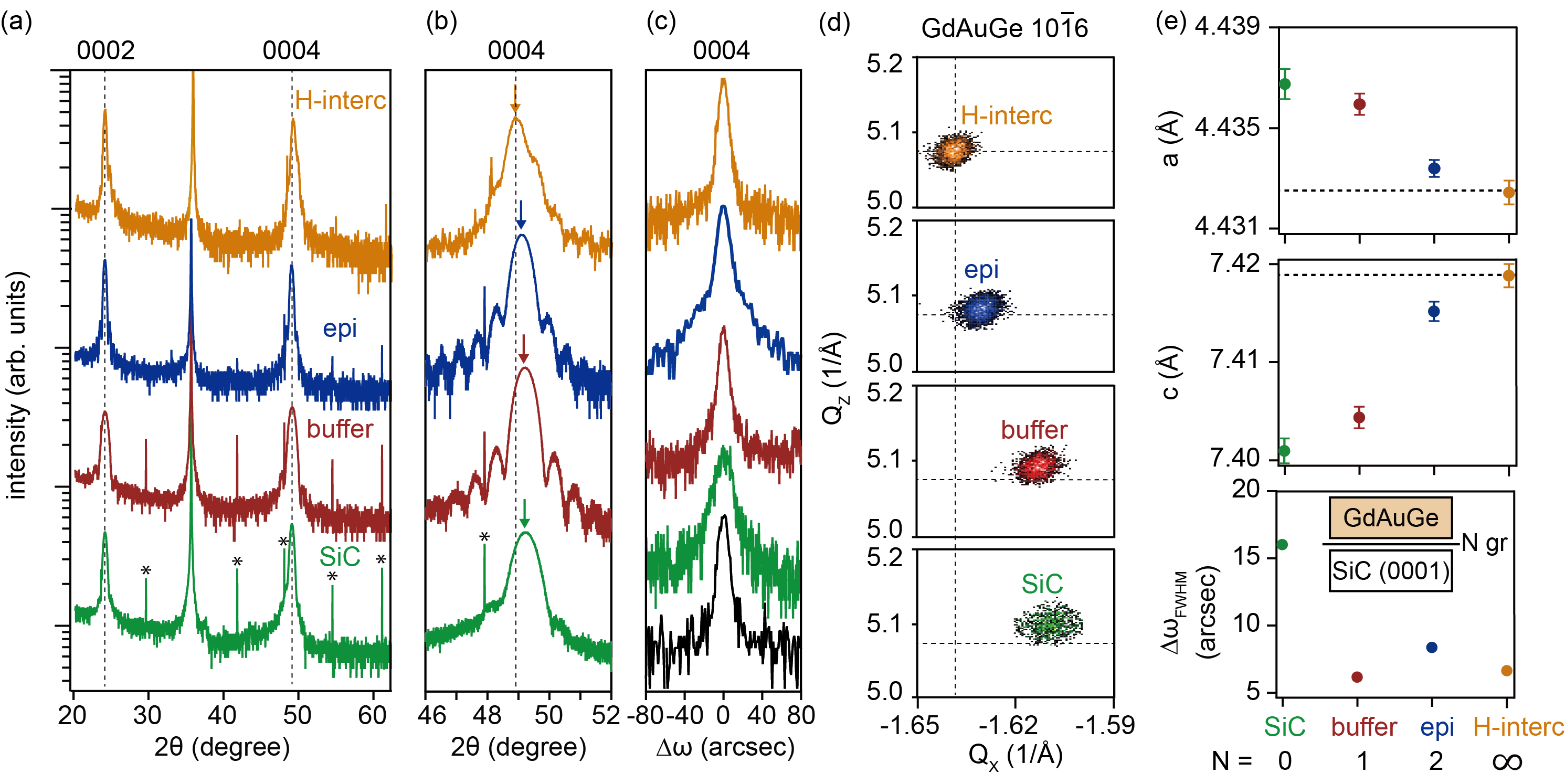}
    \caption{\textbf{X-ray diffraction of GdAuGe on $N$ graphene/SiC (0001).} (a) Wide angle symmetric $2\theta-\omega$ scans displaying only the expected $000L$ GdAuGe film reflections. SiC substrate reflections are marked by *.
    (b) Zoom in of the $0004$ reflection displaying sharp thickness fringes. Dotted line marks the position of the relaxed bulk reflection. (c) Rocking curve of films, compared to the SiC substrate (black). (d) Asymmetric reciprocal space maps. (e) Summarized $a$ and $c$ lattice parameters, and rocking curve width, as a function of $N$.
    }
    \label{supp:xrd}
\end{figure}

\begin{figure}
    \centering
    \includegraphics[width=0.8\linewidth]{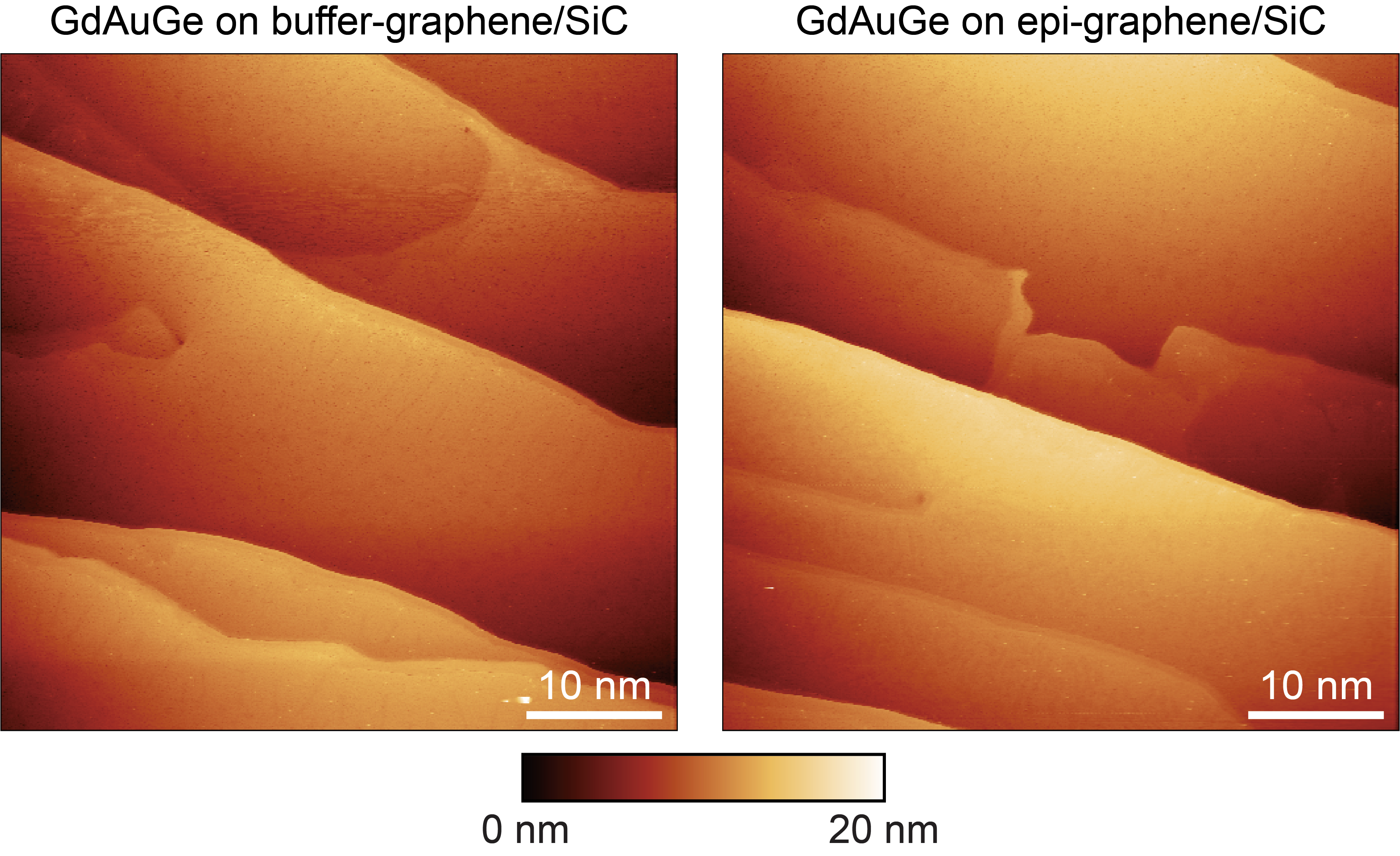}
    \caption{AFM of GdAuGe on buffer and epitaxial graphene, showing a smooth morphology that follows step terraces.}
    \label{supp:afm}
\end{figure}

\begin{figure}[h]
    \centering
    \includegraphics[width=0.9\linewidth]{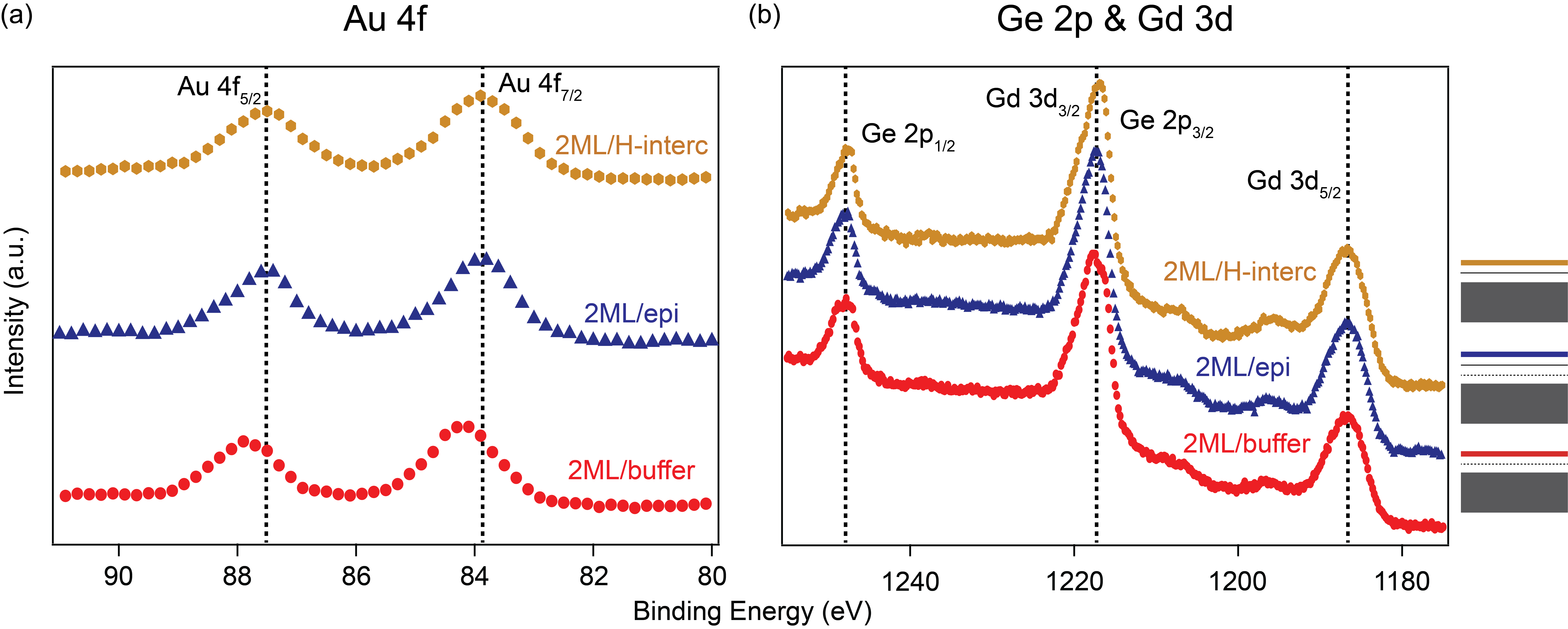}
    \caption{\textbf{In situ XPS (Mg $K\alpha$) of 2 ML GdAuGe on buffer, epitaxial, and H-intercalated graphene/SiC, showing similar chemical interactions between GdAuGe and these different forms of graphene.} (a) Au $4f$ and (b) Ge $2p$ and Gd $3d$ core level line shapes on the three surfaces are nominally identical, within resolution limits, suggesting similar chemical interactions between GdAuGe and these three forms of graphene. For the film on buffer graphene, the uniform shift of the Au $4f$ to higher binding energy suggests an Au to graphene charge transfer.}
    \label{supp:xps}
\end{figure}

\begin{figure}
    \centering
    \includegraphics[width=1\linewidth]{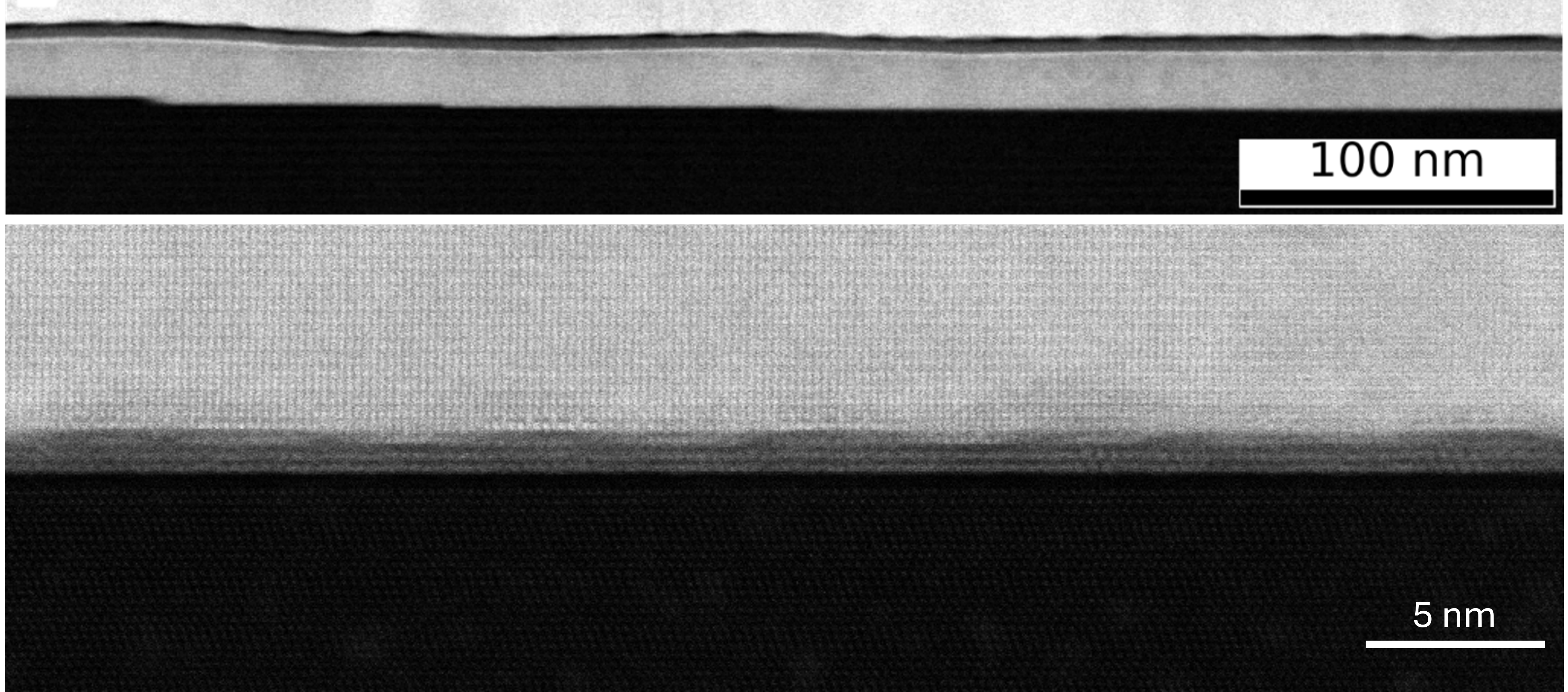}
    \caption{\textbf{Spatial extent of the finite-range GdAuGe interface on buffer graphene/SiC.} Top: low magnification HAADF-STEM image showing continuous 20 nm thick GdAuGe film. Bottom: higher magnification image showing that the 2–3-layer interfacial state persists laterally.}
    \label{supp:TEM_wide}
\end{figure}

\begin{figure*}
    \centering
    \includegraphics[width=0.95\textwidth]{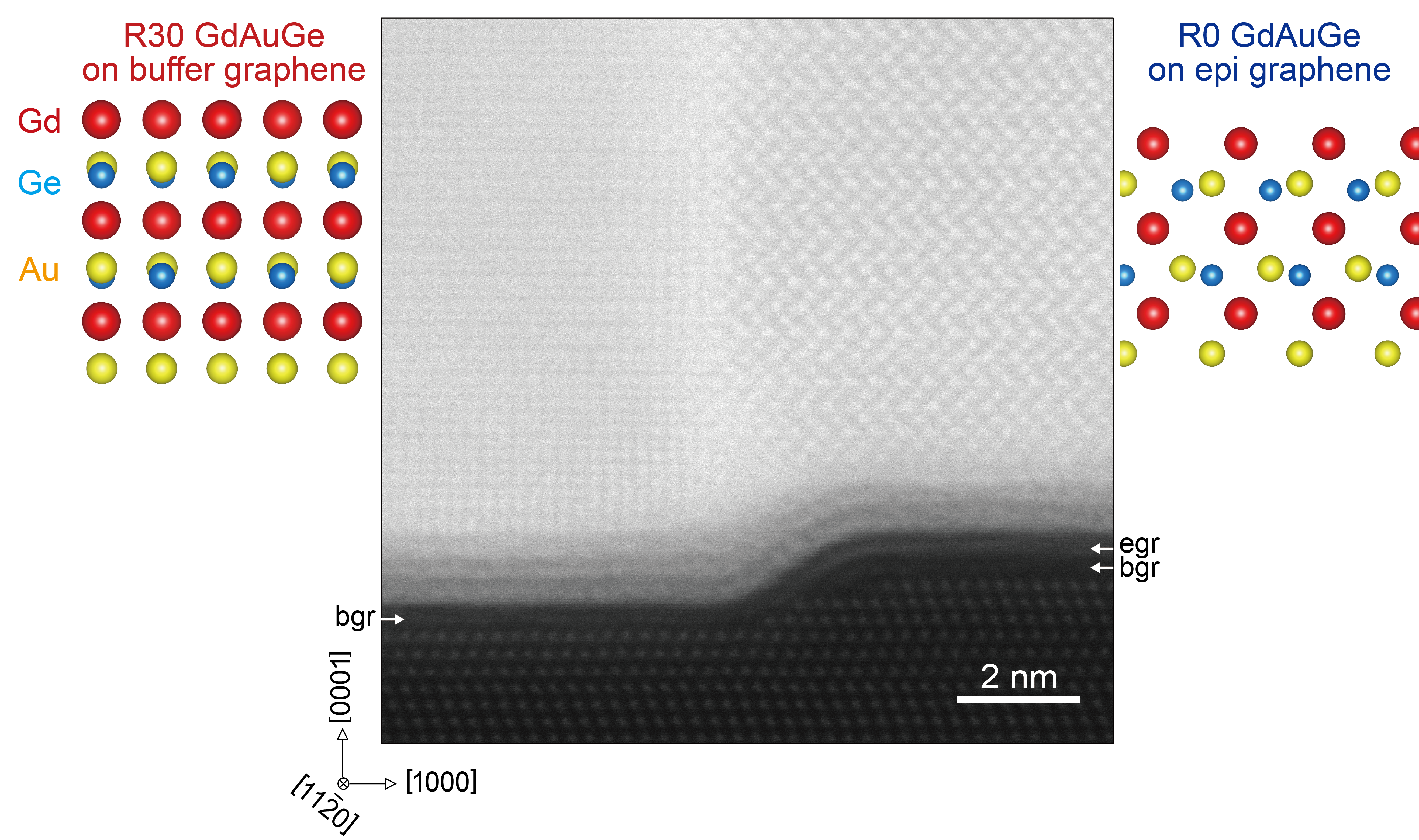}
    \caption{Cross-sectional HAADF-STEM image of GdAuGe grown across an atomic step separating adjacent buffer graphene (bgr) and epitaxial graphene (egr) regions on SiC. GdAuGe adopts the R30 orientation above buffer graphene and switches to R0 above epitaxial graphene. Because the two regions are part of the same sample and experienced identical growth conditions, the local change in epitaxial registry provides an internal control linking orientation selection to graphene thickness rather than sample to sample variations in growth conditions.}
    \label{supp:step}
\end{figure*}

\begin{figure*}
    \centering
    \includegraphics[width=0.95\textwidth]{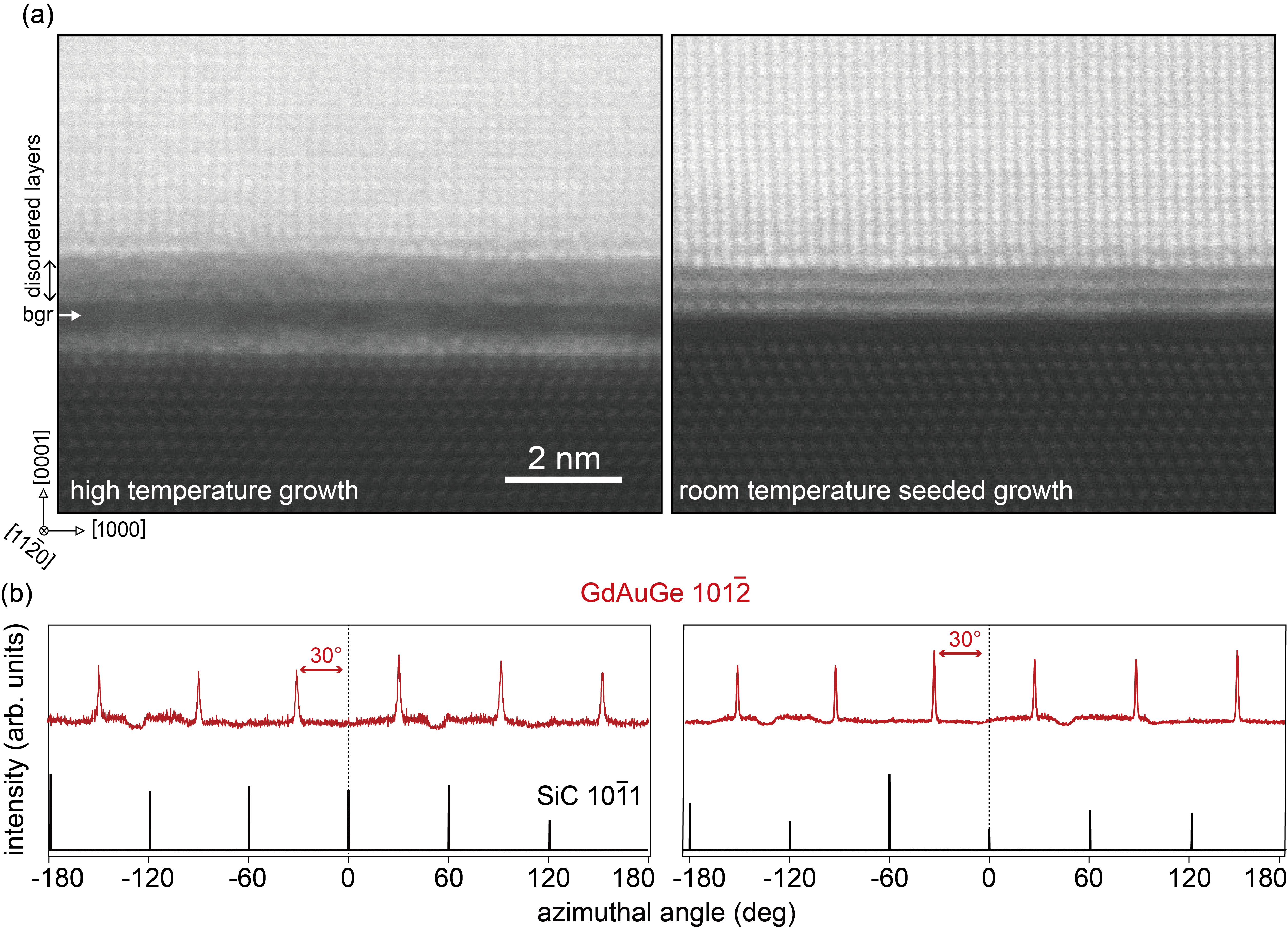}
    \caption{\textbf{Frustrated interface formation through distinct growth pathways.} (a) Cross-sectional HAADF-STEM images of GdAuGe/buffer-graphene/SiC grown directly at $450\degree$C (left) and grown using a room temperature seed followed by annealing to $450\degree$C (right). Both growth pathways produce a finite-range ordered GdAuGe interfacial layer beneath the long-range ordered film, demonstrating that formation of the frustrated interface does not require the room-temperature seed-and-anneal pathway. Greater intercalation beneath the buffer graphene (bgr) is observed for direct high temperature growth, indicating that intercalation is not required for formation of the frustrated interface. (b) X-ray azimuthal scans of the GdAuGe $10\bar{1}2$ reflection for the corresponding directly grown (left) and room-temperature-seeded (right) samples, showing the same R30 orientation selection independent of anneal pathway.}
    \label{supp:history}
\end{figure*}

\begin{figure}
    \centering
    \includegraphics[width=0.55\linewidth]{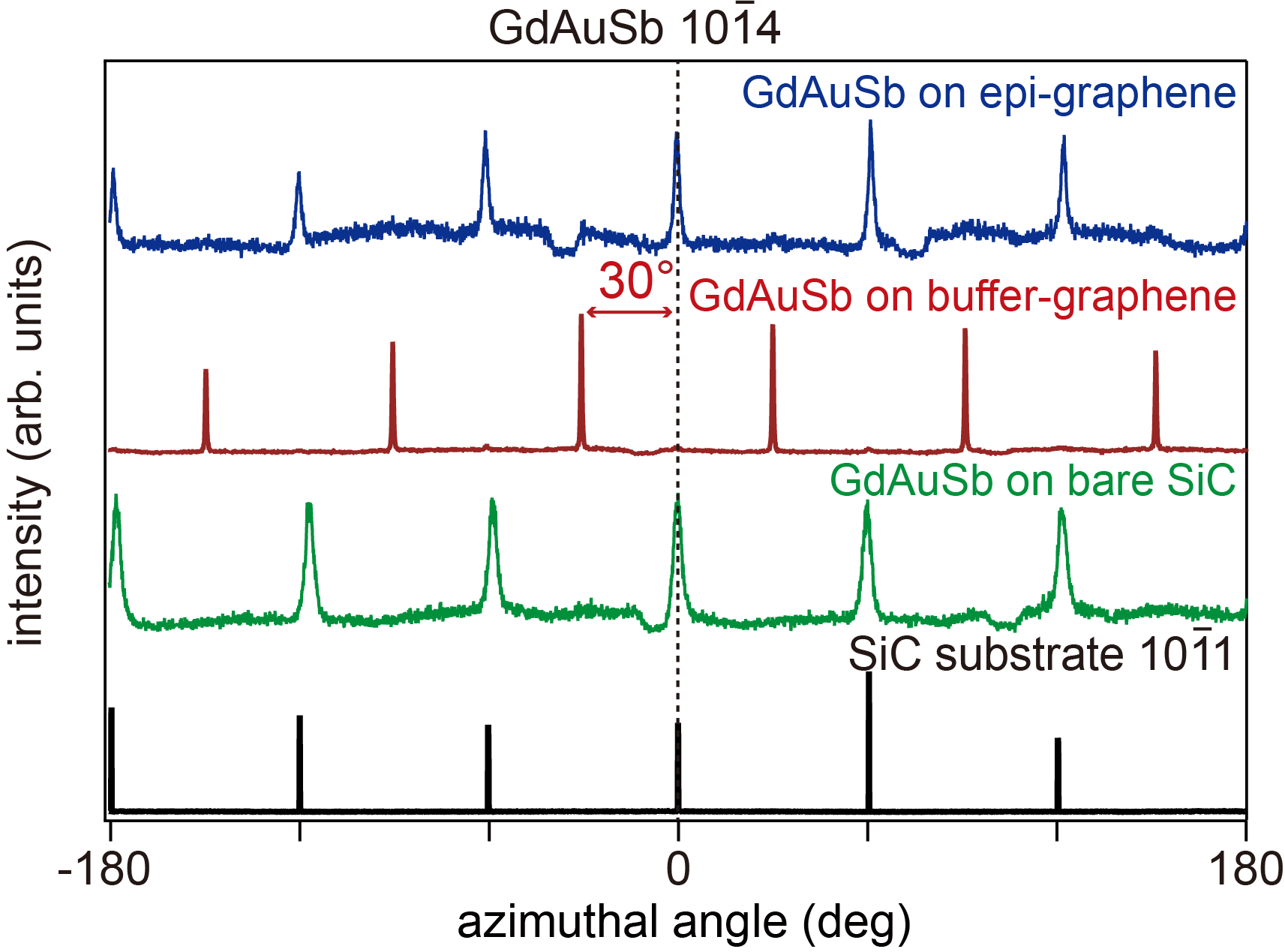}
    \caption{Graphene-dependent orientation selection in GdAuSb. Azimuthal x-ray diffraction shows the same $R0 \rightarrow R30\rightarrow R0$ sequence for GdAuSb on bare SiC, buffer graphene/SiC, and epitaxial graphene/SiC, respectively, as observed for GdAuGe.
    }
    \label{supp:GdAuSb}
\end{figure}

\begin{figure}
    \centering
    \includegraphics[width=0.3\linewidth]{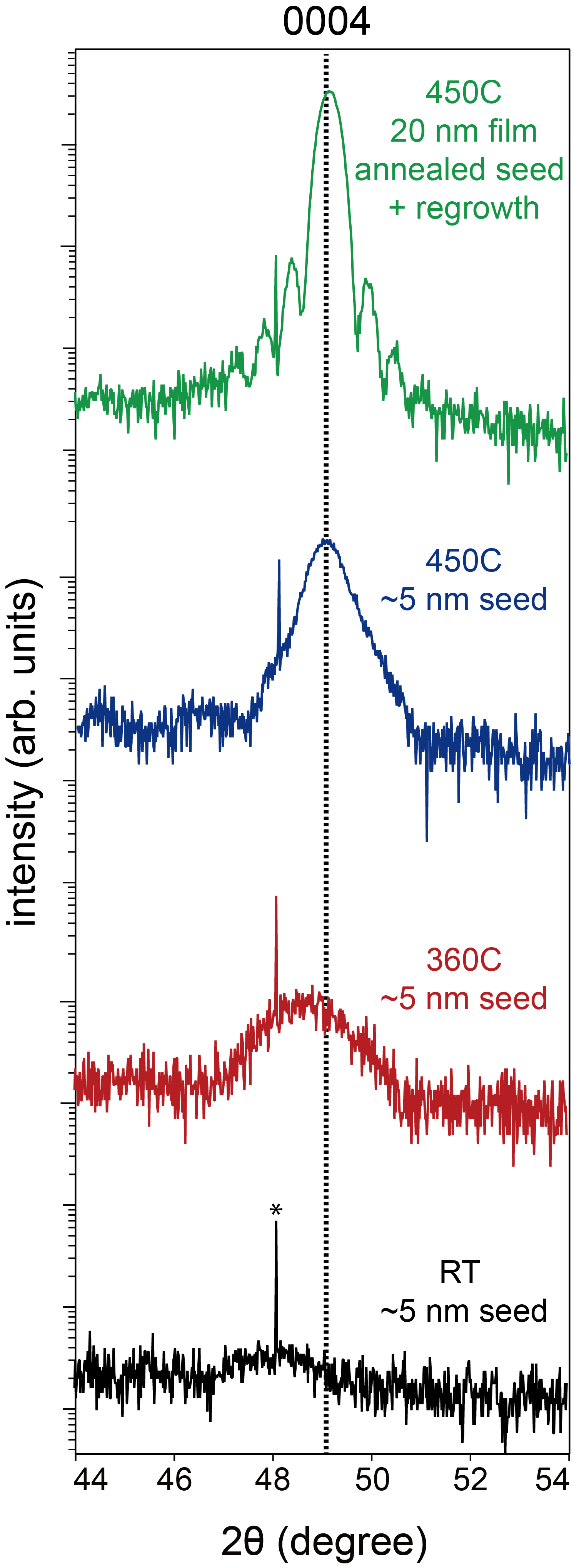}
    \caption{Anneal evolution of GdAuGe on buffer graphene/SiC measured by symmetric x-ray diffraction, tracking the evolution from room temperature amorphous (black), to strained crystal at $360
    \degree$C (red), to relaxed crystal at $450\degree$C (blue), to continued growth out to 20 nm thickness (green).
    }
    \label{supp:anneal_xrd}
\end{figure}

\begin{figure}
    \centering
    \includegraphics[width=0.6\linewidth]{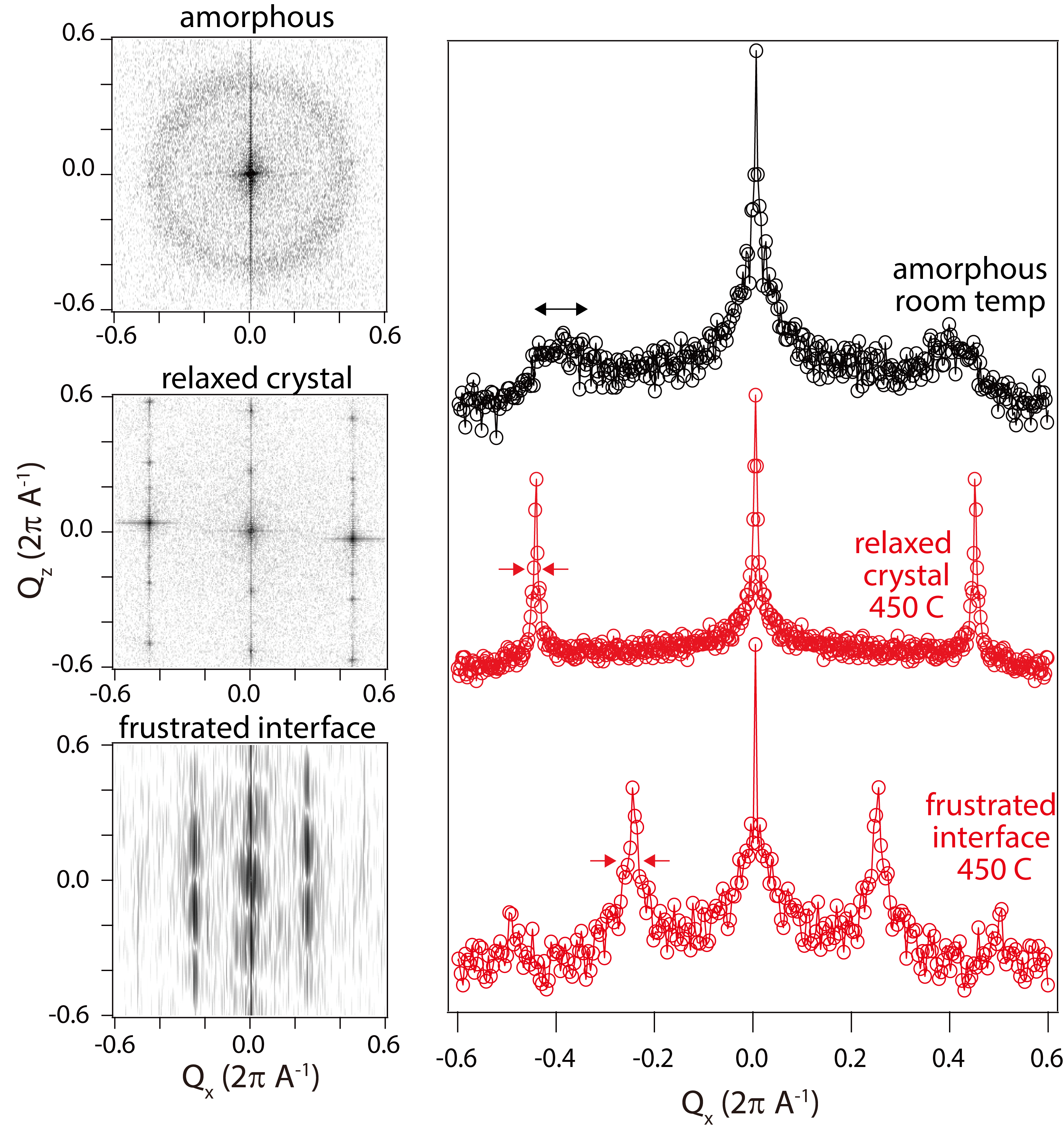}
    \caption{Extraction of coherence length $\xi=2\pi/\Delta Q_x$ from the full width at half maximum of the STEM-FFT peaks, yielding $\xi\approx10$ {\AA} for the amorphous seed, $\xi\approx30$ {\AA} for the frustrated interface, and $\xi>100$ {\AA} for the relaxed crystal. The 1D FFT line cuts are integrated from $Q_z = -0.6$ to $+0.6$ ($2\pi/$\AA).}
    \label{supp:fft_width}
\end{figure}

\begin{figure}
    \centering
    \includegraphics[width=0.4\linewidth]{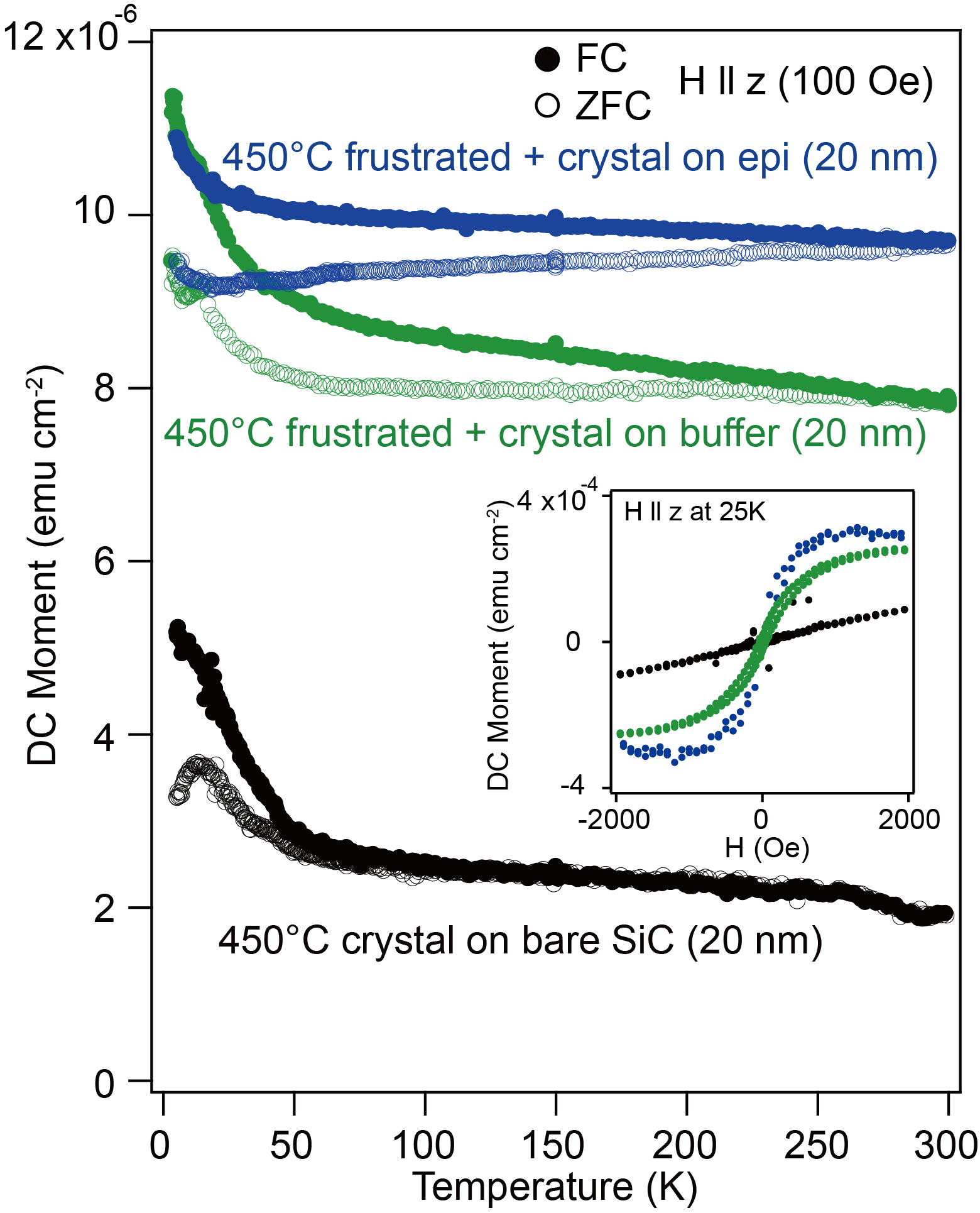}
    \caption{\textbf{Similar magnetic response of frustrated GdAuGe films on epitaxial graphene and buffer graphene.} Field cooled (FC) and zero field cooled (ZFC) magnetic moment versus temperature showing pronounced magnetic irreversibility for samples with frustrated interfaces on buffer graphene/SiC (green) and epitaxial graphene/SiC (blue), and a weaker magnetic response for crystalline GdAuGe on bare SiC (black). Inset shows $M(H)$ at 25~K.}
    \label{supp:magnet20nm}
\end{figure}

\begin{figure}
    \centering
    \includegraphics[width=.9\linewidth]{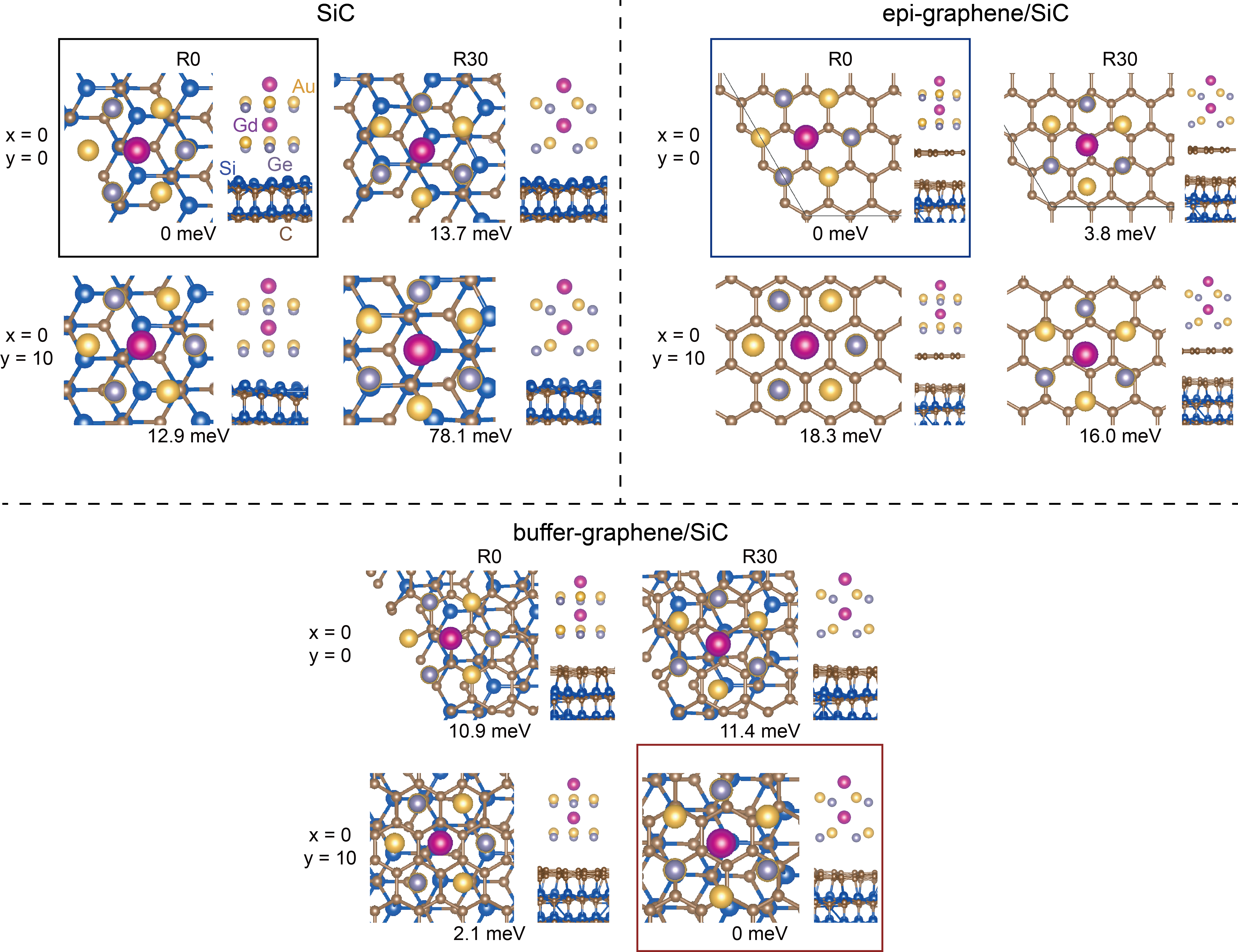}
    \caption{Slab models used for the rotated GdAuGe island formation energies $E_{form}(\theta,x,y)$ on $N$ graphene/SiC (0001) in Supplementary Fig. \ref{supp:rotation_small}. The island contains 14 atoms. The axis of rotation is through the central Gd atom. $(x,y)$ coordinates are in units of {\AA}ngstr\"{o}m. Due to computational resource limitations, the energy of the systems are calculated in a single shot, i.e. without atomic optimization.    The resulting energies are therefore used only to compare the qualitative structure of the registry-dependent energy landscape
    }
    \label{supp:rotation_model}
\end{figure}

\begin{figure}
    \centering
    \includegraphics[width=.85\linewidth]{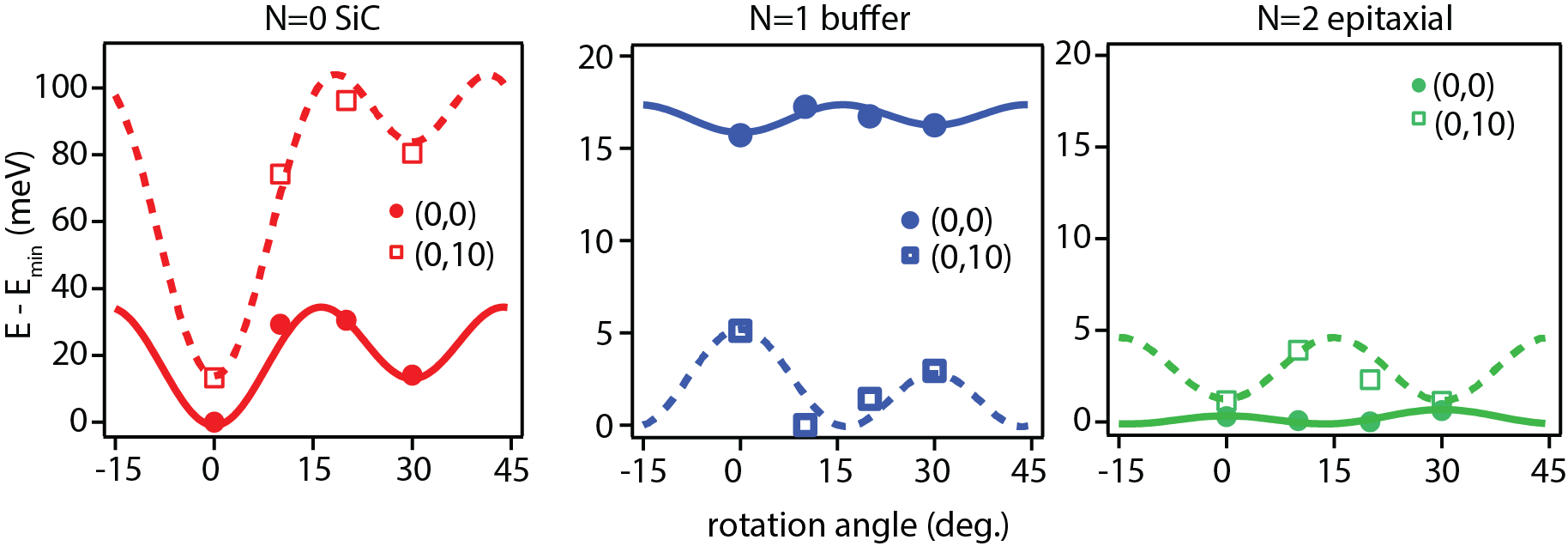}
    \caption{Formation energy of 14-atom GdAuGe island on $N$ graphene/SiC (0001), as function of in-plane rotation angle and lateral position $E_{form}(\theta,x,y)$. Energy is normalized per atom within the island. Zero angle is defined as $a_{\mathrm{GdAuGe}} \parallel a_{\mathrm{SiC}}$. The corresponding models are shown in Supplementary Fig. \ref{supp:rotation_model}. Symbols are calculated data points, continuous curves are sinusoidal guides to the eye. Multiple shallow rotational/registry-dependent minima emerge for graphene-containing surfaces, in contrast to the stronger R0 minimum on bare SiC.
    }
    \label{supp:rotation_small}
\end{figure}

\begin{figure}
    \centering
    \includegraphics[width=.85\linewidth]{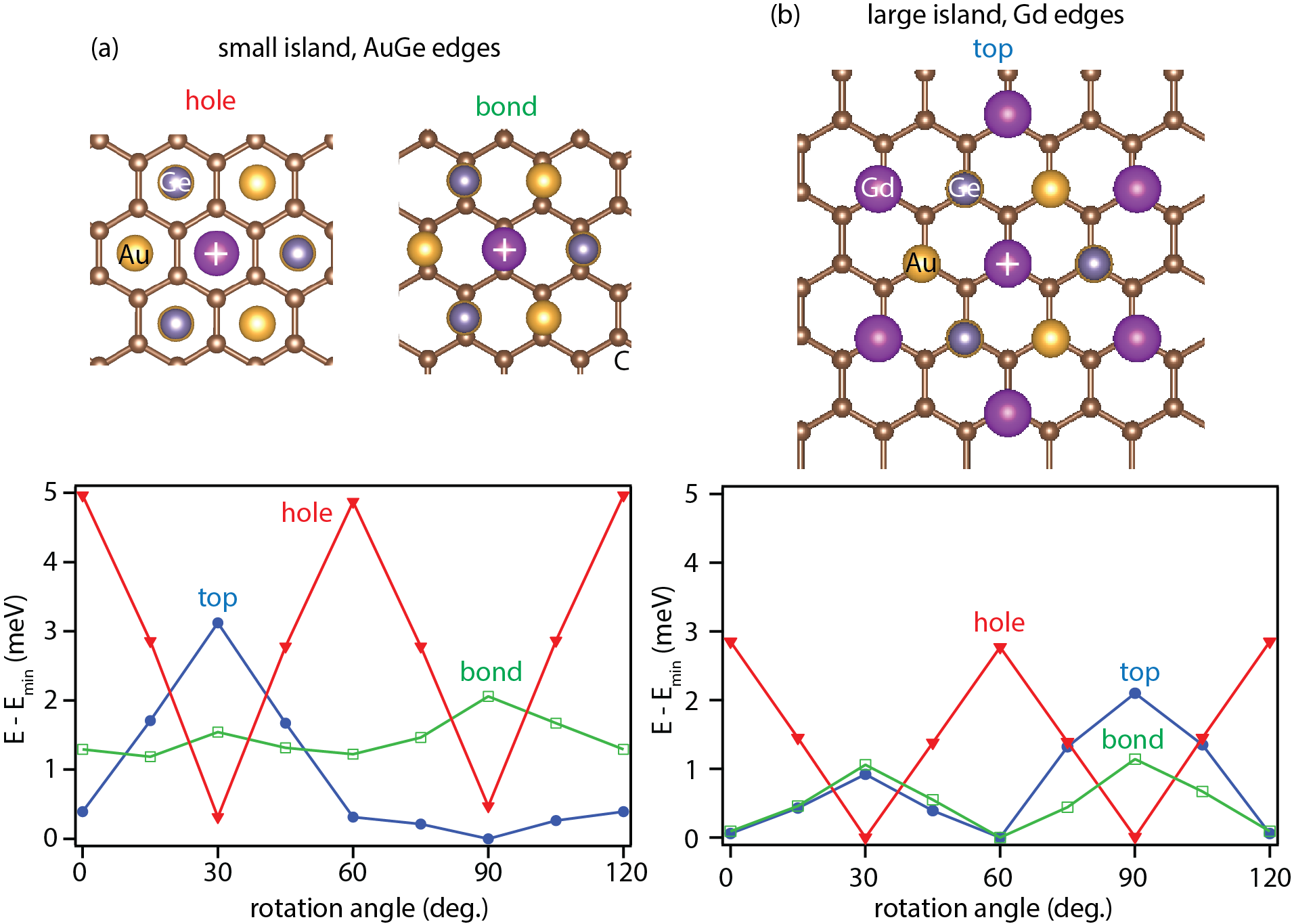}
    \caption{Island size and edge termination dependence of $E_{form}(\theta,x,y)$ for 14- and 26-atom GdAuGe islands on freestanding graphene (0001). Energy is normalized per atom within the island. ``Hole,'' ``bond,'' and ``top'' describe position of the central Gd atom (marked by a +) with respect to graphene high symmetry sites. The axis of rotation is through this central Gd atom. 
    The qualitative presence of multiple shallow minima is retained across island sizes, whereas their precise relative energies depend on island size and edge termination. This sensitivity motivates interpreting the calculations as evidence for competing configurations rather than as quantitative predictions of the experimentally selected registry.
    }
    \label{supp:rotation_gr}
\end{figure}

\end{document}